\documentclass[final]{IEEEtran}
\usepackage[dvips]{graphics}
\usepackage{epsf,rotating,color,graphicx,float,amssymb,amsmath}
\usepackage{epsfig,url,times,cite,subfigure,soul}
\usepackage{booktabs}
\usepackage{multirow,multicol}
\usepackage{lipsum}
\usepackage{algorithm,algorithmic}
\usepackage{nomencl}
\usepackage{epstopdf}

\usepackage{calc}
\newenvironment{notationlist}[1]
{\begin{list}{}%
{\renewcommand\makelabel[1]{##1\hfill}%
\settowidth\labelwidth{\makelabel{#1}}%
\setlength\leftmargin{\labelwidth+\labelsep}
\setlength\itemsep{0\baselineskip}}}%
{\end{list}}

\begin{document}

\title{\Large  Power-Traffic Coordinated Operation for Bi-Peak Shaving and Bi-Ramp Smoothing \\ --A Hierarchical Data-Driven Approach}

\author{\large Huaiguang~Jiang,~\IEEEmembership{Member,~IEEE}, Yingchen~Zhang,~\IEEEmembership{Senior Member,~IEEE}, Yuche Chen, Changhong Zhao,~\IEEEmembership{Member,~IEEE}, Jin~Tan,~\IEEEmembership{Member,~IEEE} 

{\normalsize National Renewable Energy Laboratory, Golden, CO 80401}
\vspace{-0.3in}

%%
%\thanks{H. Jiang, Y. Zhang, Y. Chen, and J. Tan are with the National Renewable Energy Laboratory, Golden, CO 80401 USA, (e-mail:yingchen.zhang@nrel.gov).
%J. Zhang and W. Gao are with the University of Denver, CO 80210 USA.}
%%
}

\date{}
\maketitle

\begin{abstract}
With the rapid adoption of distributed photovoltaics (PVs) in certain regions, issues such as lower net load valley during the day and more steep ramping of the demand after sunset start to challenge normal operations at utility companies. Urban transportation systems also have high peak congestion periods and steep ramping because of traffic patterns. We propose using the emerging electric vehicles (EVs) and the charing/discharging stations (CDSs) to coordinate the operation between power distribution system (PDS) and the urban transportation system (UTS), therefore, the operation challenges in each system can be mitigated by utilizing the flexibility of the other system.  

The proposed operation approach is designed hierarchically consists of a higher and a lower level. In the higher level, we assume integrated operation of both the PDS and UTS, and target of the operation is to minimize the social cost. Meanwhile, the target for the EVs and the CDSs as customers is to minimize their own expenditures. Then, there exists an equilibrium between two targets to determine the optimal charging/discharging price. In the lower level, the temporal \& spatial models of PDS and UTS are developed to provide a detailed analysis of the power-traffic system. Specifically, the PDS is built with a three-phase unbalanced AC power flow model, the optimal power flow (OPF) problem is relaxed with the semidefinite relaxation programming (SDP), and solved with alternating direction method of multiplier (ADMM). A dynamic user equilibrium (DUE) problem is formulated for the UTS, which is based on the static user equilibrium (SUE) with additional constraints to ensure a temporally continuous path of flow. The EVs and the CDSs function as reserves for both the PDS and UTS, and the state of charge (SOC) is considered to optimize the charging/discharging schedule and reduce the impacts to the PDS. We conducted the simulation and numerical analysis using the IEEE 8,500-bus for the PDS and the Sioux Falls system with about 10,000 cars for the UTS. Two systems are simulated jointly to demonstrate the feasibility and effectiveness of the proposed approach.
\end{abstract}


%%%%%%%%%%%%%%%%%%%%%%%%%%%%%%%%%%%%%%%%%%%%%%%%%%%%%%%%%%%%%%%%%%%%%%
\section*{Nomenclature}
\subsection*{Abbreviations}
\begin{notationlist}{MAXIMUMSPC}

\item[PDS] Power distribution system 

\item[UTS] Urban transportation system 

\item[CDS] Charging/discharging station

\item[SUE] Static user equilibrium

\item[DUE] Dynamic user equilibrium

\item[OPF] Optimal power flow 

\item[SDP] Semidefinite relaxation programming

\item[ADMM] Alternating direction method of multiplier

\item[EV] Electrical vehicle 

\item[SOC] State of charge

\end{notationlist}

\subsection*{Functions}
\begin{notationlist}{MAXIMUMSPC}

\item[$F^t_{UTY}$] Objective function of the utility and customer in time interval $t$, $t \in D_t$

\item[$F^t_{CSO}$] Objective function of the customer in time interval $t$, $t \in D_t$

\item[$F^t_{CDS}$] Objective function of the smart charging/discharging in a CDS

\item[$f^t_{PS}$] Cost function of PDS in time interval $t$

\item[$f^t_{UTS}$] Cost function of UTS in time interval $t$

\item[$f^t_{WT}$] Cost function of charging/discharging waiting time 

\item[$Q^t_{PDS}$] Netload of PDS in time interval $t$

\item[$Q^t_{DPDS}$] Total charging/discharging power of all CDS

\item[$Q^t_{DUTS}$] Total EVs for charging/discharging

\item[$F^t_{PDS}$] Objective function of OPF in PDS 

\item[$f_{k_a}(x^t_{k_a})$] Travel time function on link $k_a$ in time interval $t$, where traffic flow is $x^t_{k_a}$

\end{notationlist}

\subsection*{Parameters}
\begin{notationlist}{MAXIMUMSPC}

\item[$\mathcal{G}^{UTS}$] The graph for UTS with a node set and a link set: $\mathcal{G}^{UTS} = [\mathcal{V}^{UTS}, \mathcal{E}^{UTS}]$, where $\mathcal{V}^{UTS}$ = $\{1,2,\cdots, n_{\mathcal{V}^{UTS}}\}$, and $\mathcal{E}^{UTS} = \{1,2,\cdots, m_{\mathcal{E}^{UTS}}\}$

\item[$P_{r_us_u}$] A set of path used to connect the OD pair, which is defined as $r_u \in \mathcal{V}^{UTS}_{r_u}$ and $s_u \in \mathcal{V}^{UTS}_{s_u}$, $\mathcal{V}^{UTS}_{r_u}$ and $s_u \in \mathcal{V}^{UTS}_{s_u}$ are the original node set and destination node set, respectively

\item[$C_{k_a}$] The traffic flow capacity on link $k_a$

\item[$\mathcal{G}^{PDS}$] The graph for PDS with a node set and a link set: $\mathcal{G}^{PDS} = [\mathcal{V}^{PDS}, \mathcal{E}^{PDS}]$, where $\mathcal{V}^{PDS}$ = $\{1,2,\cdots, n_{\mathcal{V}^{PDS}}\}$, and $\mathcal{E}^{PDS} = \{1,2,\cdots, m_{\mathcal{E}^{PDS}}\}$

\item[$\underline{\Upsilon},  \overline{\Upsilon}$] The upper bound and lower bound of SOC

\item[$\gamma_1$] Weight factor between PDS and UTS

\item[$\gamma_2$] Iteration step coefficient for gradient descent

\item[$\varrho_1, \varrho_2$] The electrical power price and the congestion fee

\item[$\varrho_3, \varrho_4$] The EV parking ratios to the CDSs

\item[$\chi_{i_1}$] The capacity of CDS $i_1$
\end{notationlist}

\subsection*{Variables}
\begin{notationlist}{MAXIMUMSPC}
\item[$\theta^t_{k_a}$] The traffic flow on link $k_a$ in time interval $t$

\item[$\pi^{*t}$] The optimal charging/discharging price in time interval $t$

\item[$q^{dt}_{r_us_u}$] Number of vehicles from $r_u$ to $s_u$ departing in time interval $d$ via any path, $d \in D_t$, $D_t$ is the time interval set

\item[$h^{d_1}_{p_{r_us_u}}$] Number of vehicles assigned to path $p_{r_us_u}$ with departing time interval $d_1$

\item[$\delta^{d_1t}_{p_{r_us_uk_a}}$] A 0-1 variable to indicate in time interval $t$, whether the trip from $r_u$ to $s_u$ is assigned to path $p_{r_us_uk_a}$ via link $k_a$ departing in time interval $d_1$

\item[$V^{\phi t}_i$] Complex voltage on bus $i$ with phase $\phi $, $\phi $ $\subseteq$ $\{a,b,c\}$

\item[$I^{\phi t}_i$] Complex current on bus $i$ with phase $\phi $, $\phi $ $\subseteq$ $\{a,b,c\}$

\item[$s^{\phi t}_i$] Power injection on bus $i$, where $s^{\phi t}_i$ = $p^{\phi t}_{i}$ + $\textbf{i}q^{\phi t}_{i}$

\item[$z^{\phi }_{i}$] The complex impedance matrix $z^{\phi }_{i}$ =  $r^{\phi }_{i}$ + $\textbf{i} x^{\phi }_{i}$

\item[$S^{\phi t}_i$] The complex power from bus $i$ to bus $i^1$, where bus $i^1$ is the ancestor of bus $i$, $S^{\phi t}_i$ = $V^{\phi t}_i (I^{\phi t}_i)^H$, $H$ indicates the hermitian transpose

\item[$v^{\phi t}_{i}$, $i^{\phi t}_{i}$] $v^{\phi t}_{i}$ = $V^{\phi t}_{i} (V^{\phi t}_{i})^H$, $l^{\phi t}_{i}$ = $I^{\phi t}_{i} (I^{\phi t}_{i})^H$

\item[$x^{t,k}, y^{t, k}, \lambda^{t, k}$] The $k$th iterative variables of ADMM for distributed OPF computation

\item[$C^{t_1}_{ev, i_3}$] The discharging (positive) and charging (negative) speed of EV $i_3$ at time $t_1$
	
\end{notationlist}
% % % % % % % % % % % % % % % % % % % % % % % % % % % % % % % % % % % %
\section{Introduction}\label{sec:introduction}
\begin{figure*}[!t]
	\begin{center}
		\includegraphics[width=1.90\columnwidth, angle=0]{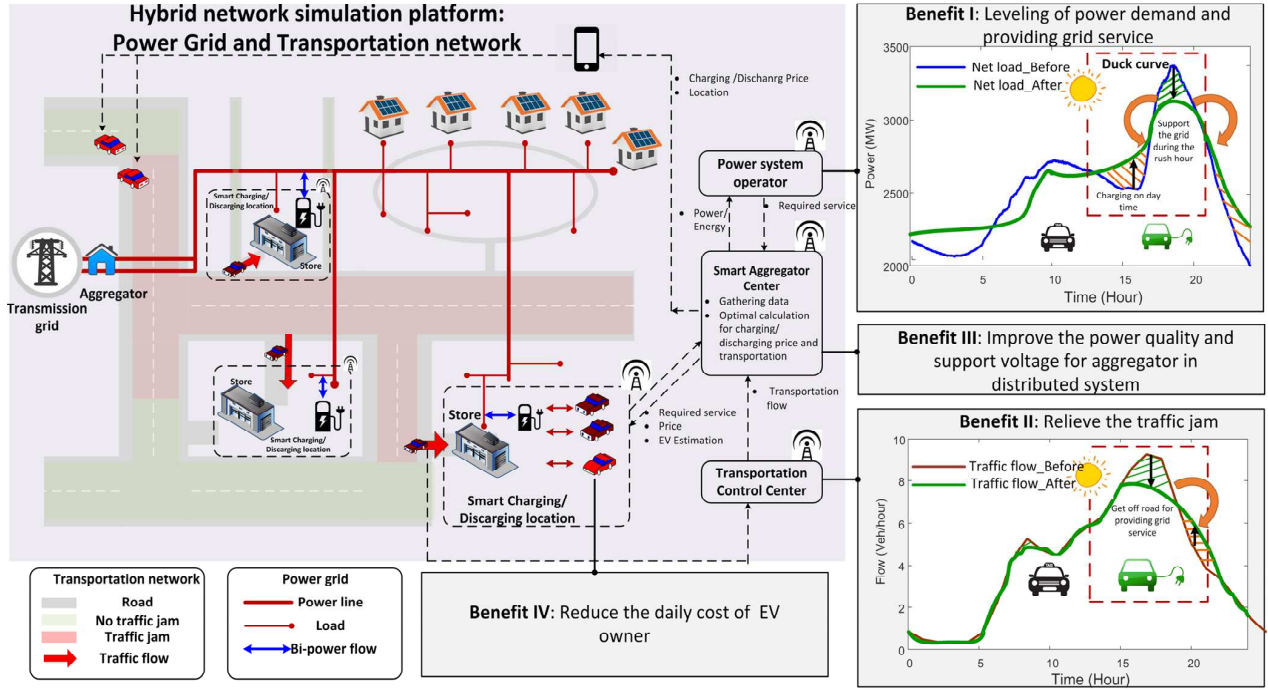}
		\caption{The main idea and the bi-peak and bi-ramp problems in a power distribution system and a transportation system in a urban area.}\label{fig:idea1}
	\end{center}
\end{figure*}

Rooftop photovaltaic (PV) gained a foothold in many power systems because of its continuously declining cost~\cite{chen12g2015seeds,cui2015sol12ar,gu2016kno12wledge}. It brings many benefit to costumers such as reduced energy cost; however, it also presents unprecedented challenges to power systems operation and control. The upper right-hand plot in Fig.~\ref{fig:idea1} shows the so-called ``duck curve", which has been recorded in some high-distribution PV regions such as California~\cite{katsigiann123is2010novel, tian2013re1view, connolly2010review}. The peak solar generation in the middle of the day sinks the netload to lower valley and then when the peak load occurs right after sunset, a huge volume of energy demand ramps up in a short time frame. This creates the artificial peak and ramping that are costly to balance using the current power system assets. 

It is not surprising that other man-made complex systems also suffer similar issues. As shown in the lower right-hand plot in Fig.~\ref{fig:idea1}, the transportation system also has high peaks and steep ramps due to fairly constrained traffic patterns at the rush hours in urban areas~\cite{gonzal65es2013evening,cheng2015urb1an,huang2017para123meterized}. Thanks to the spawn of EVs and the widely located CDSs, the two originally independent systems: the PDS and the UTS can be coupled together.

Specifically, the increasing number of EVs can be seen as a significant amount of distributed and highly manipulatable small reserves that can be used to provide demand response, enhance the system reliability, and provide other services for the power systems~\cite{fol1ey2013impacts, liu2012asse234ssment, wu2012loa234d,sortomme2012opt234imal,synchro2015jiangf234s,heymans2014econo123mic,geng2017learn89ing}. In these studies, the geographical information and transportation information are ignored, because the EVs are treated as the aggregated loads/reserves. In~\cite{sadeghi2014opt234imal}, the optimal placement of the charging stations is studied with the geographical information and transportation information. In~\cite{he2013op234timal}, based on locational marginal pricing (LMP), an optimal deployment of charging stations is proposed for EVs. In~\cite{wei2017opt456imal}, an optimal traffic-power flow is proposed with the wireless charging technology and congestion toll manipulation.

In this paper, a two-step power-traffic coordinated operation approach is proposed to address the bi-peak and bi-ramp problems for both PDS and UTS. It also provides a multifunction platform for the researches such as emission reduction, multisystem integration, and future city planning. The \textbf{main contributions} of this paper are:
\begin{enumerate}
	\item Considering the complexity of the power-traffic system, a two-step coordinate approach is designed from the higher level to lower level to operate the system in spatial and temporal domain hierarchically. In the higher level, both the PDS and the UTS are regulated and treated together as an utility to minimize the social cost. The EVs and CDSs are treated as customers to minimize the expenditure. Then, an equilibrium exists between the utility and the customers to determine the operation variables such as optimal charging/discharging price, total electrical demand, and total available EVs. In the lower level, the detailed models of PDS, UTS, EVs \& CDS are considered to specifically determine these variables in spatial and temporal domains.
	
	\item In the lower level, the PDS is built with the branch flow model, which is equivalent to the classic bus-injection model and more suitable for the PDS with radial topology~\cite{suh20va09lnc}.  There are several developed approaches for solving an optimization problem associated with a PDS, usually called OPF such as DC OPF~\cite{st456ott2009dc}, genetic algorithm~\cite{b45tzs2002optil,yigu2014statistical1}, and OPF relaxation with the second-order cone program (SOCP)~\cite{pe78ng2014distributed}. Based on~\cite{pe015dis456ed,dall2013di23stributed}, we relax an OPF with the SDP for the three-phase unbalanced PDS. Meanwhile, several distributed algorithm are designed to solve the OPF such as dual decomposition~\cite{lam212dti123bud} and predictor corrector proximal multiplier (PCPM)~\cite{li2012dem45and}. Based on~\cite{pe015dis456ed,boyd2011d45istributed}, the ADMM is applied to decompose the OPF problem and reduce computation time.
	
	\item In the lower level, for the UTS, the DUE is widely used to estimate the traffic volume, congestion time, and travel cost for each link~\cite{pat015fic}. Based on the assumption that all travelers can choose the best route to minimize the cost of transportation, in the Wardrop user equilibrium (UE), the travel times of all paths are equal for each original-destination (OD) pair and less than any unused paths~\cite{jia2056514nk,sheffy1985ur123ban}. In~\cite{wei2017opt456imal}, a static user equilibrium assignment is used to integrate with the PDS. Considering that the EV behaviors can be impacted by the charging/discharging prices, a DUE is applied to keep the temporal domain continuous for the UTS.
	
	\item In this paper, the EVs and the CDSs are designed as the ``reserves" for both PDS and UTS, respectively. Considering the SOC~\cite{he2013n1ew,pattipati2011sy11stem}, an optimal charging/discharging schedule is designed to meet the requirements of PDS and UTS and reduce the impacts to PDS. The test bench consists of real systems, the PDS is the IEEE 8,500-bus distribution system, and the UTS is the Sioux Falls transportation system with about 10,000 EVs, to demonstrate the proposed approach.
\end{enumerate}

 Based on Fig.~\ref{fig:idea1}, we acknowledge that cybersecurity is a critical aspect to consider in the proposed power-traffic system operation. Also, the large volume of data generated by the proposed power-traffic system requires a high-speed, flexible and reliable communication network. Because the proposed power-traffic system contains two complex systems, many industrial communication network infrastructures for PDS and UTS operations need to be attached, such as 3G, 4G, and WiFi~\cite{sed12jelmaci2017hierarchical}. The multi-network integration brings a lot of challenge for the 1). real-time monitoring \& anomaly detection~\cite{jiang2014fault,jiang2016spatial1}, 2). data transmission, storage, \& security, 3). attack analysis \& mitigation~\cite{ten2010cyber12security, islam2012wirel12ess,geng2015da12ta,jiang2017big}. In this paper, we assume that all the messages such as electrical power price, traffic congestion information, and various control signals can be transmitted correctly in real-time without any cybersecurity issues. 

The paper is organized as follows. In Section~\ref{sec:formulation}, the flowchart of the proposed approach is introduced. In Section~\ref{Sec:Equilibrium}, the equilibrium between the utility and the customers is designed to determine the optimal system variables. In Section~\ref{Sec:UTS}, based on SUE, a DUE is applied to model the UTS and compute the dynamic traffic flow. In Section~\ref{sec:PDS}, based on the branch flow model, the OPF problem is relaxed with SDP and solved with ADMM.  In Section~\ref{Sec:EVs}, considering the SOC and the EVs behaviors, an optimal charging/discharging schedule is designed to reduce the impacts to the PDS. In Section~\ref{sec:Implementation}, the numerical results are presented to validate the proposed approach. The conclusion is presented in Section~\ref{Sec:Concl}.

% % % % % % % % % % % % % % % % % % % % % % % % % % % % % % % % % % % % %
\section{The Flowchart of the Proposed Approach}\label{sec:formulation}
\begin{figure*}[t!]
	\begin{center}
		\includegraphics [width=1.25\columnwidth, angle=90]{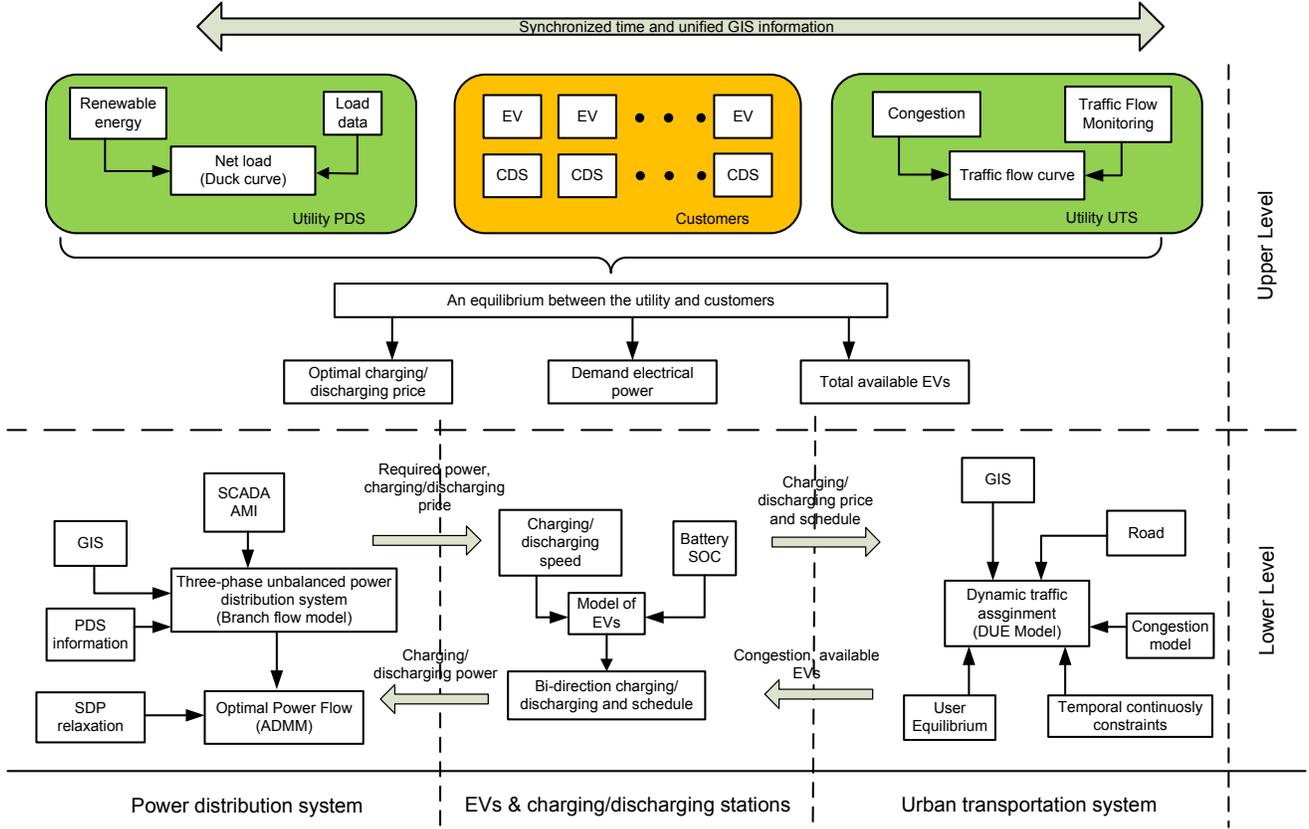}
		\caption{Flowchart of the proposed power-traffic coordinate operation approach.}\label{fig:flowchart}
	\end{center}
\end{figure*}

The proposed two-step power-traffic operation approach is shown in Fig.~\ref{fig:flowchart}. In the higher level, the power-traffic system consists of two major parts: the utility part (taking the PDS and the UTS as a whole) shown in green, and customer part (taking the CDSs and related EVs) shown in yellow. At time interval $t$, the duck curve of the netload is recorded with the renewable energy generation data and end user load data in PDS. Meanwhile, the travel data is collected with the traffic monitoring data and congestion model in UTS. Then, as shown in the two green blocks, the PDS and UTS are treated as a regulated utility to minimize the social cost. In the yellow block, the EVs and CDSs are treated as the customers to minimize their expenditures. There exists an equilibrium to determine the system operation variables such as optimal charging/discharging prices, total demand electrical power, and total required EVs. These variables are fitted into the lower level as inputs.

In the lower level, the power-traffic system consists of three major parts: the PDS part, the EVs \& CDSs part, and the UTS part, which determine the system operation variables in spatial and temporal domains with detailed models. First, in the UTS, to accurately simulate the dynamic traffic system in spatial and temporal domain, the DUE is built based on SUE with temporal continuous constraints. The proposed DUE problem can be transfered into a convex problem and solved with optimal results. Second, in the PDS, the objective function $F_{PDS}$ is designed to minimize the cost of the PDS, which is built with a three-phase unbalanced branch flow model. Based on the SDP relaxation, $F_{PDS}$ can be minimized in an OPF problem, which is solved with ADMM. Third, with the specified electrical power demands and available EVs for each CDS, an objective function $F_{CDS}$ is designed to minimize the charging/discharging impacts to the PDS while considering the SOC.

In addition, in this paper, it is assumed that all the drivers know the traffic information very well, and can get the optimal charing/discharging price through the wireless communication immediately. The user-equilibrium in UTS network can be reached in a very short time.

\section{Equilibrium Design Between Utilities and Customers (Higher Level)}\label{Sec:Equilibrium}

\subsection{Equilibrium Design}
In this paper, we focus on a small city with tens of thousands of cars, which means any single user behavior has a very weak impact on the equilibrium design in the higher level. The utility consists of PDS and UTS. A customer consists of a CDS and related EVs.

\textbf{Utility Objective (Minimize Social Cost)}
\begin{figure*}[!t]
	%\scriptsize
	%\setcounter{equation}{15}
	\begin{equation}\label{equ:ISO}
		%\begin{align}
		F^t_{UTY} = \min_{P^t_{i_1}, N^t_{i_1}} \bigg\{ f^t_{PS} (Q^t_{PDS} - Q^t_{DPDS} (P^t_{i_1})) 
		  + \gamma_1 f^t_{UTS}(Q^t_{UTS} - Q^t_{DUTS} (N^t_{i_1})) + f^t_{WT}(\tau_1 Q^t_{DUTS} (N^t_{i_1})) \bigg\}
		%\end{align}
	\end{equation}
	%\hrulefill
	\vspace*{0.0in}
\end{figure*} 

As shown in (\ref{equ:ISO}), $F^t_{UTY}$ is the objective function of the utility, which is designed to minimize the social cost with the variables $P^t_{i_1}$ and $N^t_{i_1}$, $t$ is a time interval $t\in D_t$. $P^t_{i_1}$ is the electrical power charging (negative) or discharging (positive) at CDS $i_1$. $\gamma_1$ is the weight coefficient of UTS. $N^t_{i_1}$ is the number of available EVs on CDS $i_1$. 
The constraints of (\ref{equ:ISO}) are illustrated as follows:
\begin{subequations}\label{equ:ISOcon}
	\begin{align}
	& \underline{P^t_{i_1}}\leq P^t_{i_1} \leq \overline{P^t_{i_1}},  \label{equ:ISO1} \\
	& 0 \leq \gamma_1 \leq \overline{\gamma_1} \label{equ:ISO2} \\
	& 0 \leq N^t_{i_1} \leq \overline{N^t_{i_1}} \label{equ:ISO3}
	\end{align}
\end{subequations}

$f^t_{PS}$ is defined as a convex function~\cite{park2005particle}, which determine the PDS cost in time interval $t$,  $Q^t_{PDS}$ is the netload of PDS, and $Q^t_{DPDS} (P^t_{i_1})$ is the total charging/discharging power of all CDSs, which can be defined as

\begin{equation}
Q^t_{DPDS} (P^t_{i_1}) = \sum_{i_1 \in N_{CDS}}P^t_{i_1} \label{equ:EV1}
\end{equation}
where $N_{CDS}$ is a set of CDS. $f^t_{UTS}$ is defined as a convex function, which determines the UTS cost in time interval $t$, $Q^t_{UTS}$ is the traffic load, and $Q^t_{DUTS} (N^t_{i_1})$ is the total EVs for charging/discharging. $f^t_{WT}(\tau_1)$ is the time cost with charging/discharging time $\tau_1$. Considering the number of EVs is very large, $Q^t_{DUTS} (N^t_{i_1})$ can be estimated as

\begin{equation}
Q^t_{DUTS} (N^t_{i_1}) = \sum_{i_1 \in N_{CDS}}N^t_{i_1} = \sum_{i_1 \in N_{CDS}}\frac{P^t_{i_1}}{C^t_1} \label{equ:EV2}
\end{equation}
where $C^t_1$ is the average charging/discharging speed of all the EVs. $f^t_{WT}$ is the waiting cost when an EV is parking for charging/discharging, which is also defined as a convex function. 
In summary, the proposed utility objective function $F^t_{UTY}$ is convex, which indicate an optimal point existed.

\textbf{Customer Objective (Minimize Own Expenditure)}
Based on (\ref{equ:EV2}), the objective function of a customer (a CDS and related EVs) can be designed as follows:
\begin{subequations}\label{equ:obj_cus}
	\begin{align}
	& F^t_{CSO} = \min_{P^t_{i_1}}f^t_{WT}(\tau_1\sum_{i_1} \frac{P^t_{i_1}}{C^t_1}) - \pi^t P^t_{i_1} \label{equ:cus1}\\
	& s.t.\ \ (\ref{equ:ISO1}),  \underline{\pi^t}\leq \pi^t \leq \overline{\pi^t} \label{equ:cus2}
	\end{align}
\end{subequations}
where the objective function of the customer consists of two parts: the time consumption in the CDSs and the benefit/cost in charging/discharging. $\pi^t$ is the electrical power price, and $\pi^t P^t_{i_1}$ is the benefit (by discharging, positive) or cost (by charging, negative) at CDS $i_1$. As discuss above, the objective function of the customer $F^t_{CSO}$ is also convex.

\subsection{Solutions}
An equilibrium $\pi^{*t}$ and $P^{*t}_{i_1}$ exists, and the optimal price $\pi^{*t}$ can be computed as~\cite{boyd2004conv456ex,li2011opt789imal}
\begin{equation}\label{equ:opti}
%\scriptsize
%\begin{align}
\pi^{*t} = f'^t_{PS}(Q^t_{PDS} - \sum_{i_1}P^{*t}_{i_1}) + \frac{\gamma_1}{C^t_1} f'^t_{UTS}(Q^t_{UTS} - \sum_{i_1} \frac{P^{*t}_{i_1}}{C^t_1})
%\end{align}
\end{equation}  
where the optimal price depends on the PDS part $f'^t_{PS}$ and the UTS part $f'^t_{UTS}$. Based on the gradient descent, a distributed algorithm is proposed to optimize the utility and the customers jointly the equilibrium. Specifically, the objective function of each customer can be computed independently to the optimal price $\pi^{*t}$.
At $k_1$-th iteration:

\begin{subequations}\label{equ:equ_iter}
	%\scriptsize
	\begin{align}
	& \pi^{k_1,t} = \nonumber\\
	& f'^t_{PS}\bigg(Q^{t}_{PDS} - \sum_{i_1}P^{k_1,t}_{i_1}\bigg) + \frac{\gamma^1_1}{C^t_1} f'^t_{UTS}\bigg(Q^{t}_{UTS} - \sum_{i_1} \frac{P^{k_1,t}_{i_1}}{C^t_1}\bigg) \label{equ:priceup}\\
	& P^{k_1+1, t}_{i_1} = P^{k_1, t}_{i_1} + \gamma_2\bigg(\frac{\tau_1}{C^t_1}f'^t_{WT}(\tau_1\sum_{i_1}\frac{P^t_{i_1}}{C^t_1})-\pi^{k_1,t}\bigg) \label{equ:powerup}\\
	& (\pi^{k_1,t}, P^{k_1+1, t}_{i_1}) = [\overline{\pi^{k_1,t}}, \overline{P^{k_1+1, t}_{i_1}}]^{\varpi}
	\end{align}
\end{subequations}
where the optimization stepsize can be set as $0 \leq \gamma_2$, and the algorithm can converge when $\gamma_2$ is small enough~\cite{bert123sekas1989parallel}. $[\cdot]^{\varpi}$ means that the results are projected onto the set $\varpi$ defined by (\ref{equ:cus2}).

In summary, in this section, the optimal price $\pi^{*t}$ can be computed in the higher level, and the optimal results $P^{*t}_{i_1}$ gives the upper bound of power injection in (\ref{equ:constain_PDS_s}). In the following sections, the models of the UTS, PDS, and CDS are built to provide detailed information in both temporal and spatial domains about how to operate the power-traffic system in the lower level.

\section{Dynamic User Equilibrium Assignment for UTS (Lower Level) }\label{Sec:UTS}
Compared with the SUE, the DUE can accurately describe the traffic dynamics in a set of successive time intervals. In this paper, the DUE is based on the SUE with a temporal generalization~\cite{ja123nson1991dynamic,sheffy1985ur123ban}. 
\subsection{Basic Model of UTS}\label{Sec:UTS1}
The UTS can be represented in a graph with a node set and a link set: $\mathcal{G}^{UTS} = [\mathcal{V}^{UTS}, \mathcal{E}^{UTS}]$, where $\mathcal{V}^{UTS}$ = $\{1,2,\cdots, n_{\mathcal{V}^{UTS}}\}$, and $\mathcal{E}^{UTS}$ = $\{1,2,\cdots, m_{\mathcal{E}^{UTS}}\}$. The origin and destination (OD) pair can be defined as $r_u \in \mathcal{V}^{UTS}_{r_u}$ and $s_u \in \mathcal{V}^{UTS}_{s_u}$, a set of paths $P_{r_us_u}$ is used to connect the OD pair, and the traffic flow between the OD pair can be represented as $q^d_{r_us_u}$ with departing time $d_t \in D_t$, where $D_t$ is a set of all time intervals. Then, the congestion model can be presented as follows~\cite{manual1964b123ureau}:

\begin{equation}\label{equ:cong}
f_{k_a}(\theta^t_{k_a}) = f^0_{k_a}\bigg[1+0.15(\frac{\theta^t_{k_a}}{C_{k_a}})^4\bigg]
\end{equation}
where $f_{k_a}(\theta^t_{k_a})$ is the travel time (travel impedance) when the traffic flow at link $k_a \in K_A$ is $\theta^t_{k_a}$ in time interval $t\in D_t$, $f^0_k$ is the initial travel time, and $C_{k_a}$ is the traffic flow capacity.
\subsection{Dynamic User Equilibrium}
Based on the SUE, the DUE can be built as follows~\cite{ja123nson1991dynamic}:

\begin{subequations}\label{equ:due_equa}
	\begin{align}
	& J_{DUE} = \min_{k_a, \theta^t_{k_a}} \sum_{k_a\in K_A}\sum_{t\in D_t} \int\limits^{\theta^t_{k_a}}_{0}f^t_{k_a}(w)dw  \label{equ: DUE1} \\
	s.t.\ \  & \theta^t_{k_a} = \sum_{p_{r_us_u}\in P_{r_us_u}}\sum_{d_1\in D_t} h^{d_1}_{p_{r_us_u}} \delta^{d_1t}_{p_{r_us_uk_a}}  \label{equ:DUE2} \\
	& q^{d_1}_{r_us_u} = \sum_{p_{r_us_u}\in P_{r_us_u}} h^{d_1}_{p_{r_us_u}}  \label{equ:DUE3} \\
	& b^{d_1}_{p_{r_us_u}} = \sum_{t\in D_t}\sum_{k_a \in K_a}f^t_{k_a}(\theta^t_{k_a})\delta^{d_1t}_{p_{r_us_u}k_a} \label{equ:DUE4} 
	\end{align}
\end{subequations}
where $h^{d_1}_{p_{r_us_u}}$ is the number of vehicles in path $p_{r_us_u}$ with departing time interval $d_1$, $b^{d_1}_{p_{r_us_u}}$ is the travel time from $r_u$ to $s_u$ via path $p_{r_us_u}$, and $\delta^{d_1t}_{p_{r_us_uk_a}}$ is a 0-1 variable and can be defined as

\begin{equation}\label{equ:OSS01}
%\scriptsize
%\begin{align}
\delta^{d_1t}_{p_{r_us_u}k_a} =\left\{ \begin{array}{ll}
1    & \mbox{If in time interval $t$, link $k_a$ is in } \\
& \mbox{path $p_{r_us_u}$ with departing time $d_1$}  \\
0    & \mbox{otherwise}
\end{array}
\right .
%\end{align}
\end{equation}

In time interval $t$, the traffic flow $\theta^t_{k_a}$ on link $k_a$ is given as in (\ref{equ:DUE2}), which equals to the sum of the traffic flows $h^{d_1}_{p_{r_us_u}} \geq 0$ via link $k_a$ for any departing time interval $t \in D_t$ and any path $p_{r_us_u}\in P_{r_us_u}$. The number of vehicles from  $r_u$ to $s_u$ is given as in (\ref{equ:DUE3}), which equals to the sum of traffic flows $h^{d_1}_{p_{r_us_u}}$ on any assigned path $p_{r_us_u}\in P_{r_us_u}$. The total time consumption of path $p_{r_us_u}$ from $r_u$ to $s_u$ is given as in (\ref{equ:DUE4}), which keeps the temporal continuous traffic flow for all time intervals $t\in D_t$ with a temporal unique constraint:

\begin{equation}\label{equ:sum}
\sum_{t\in D_t}\delta^{d_1t}_{p_{r_us_u}k_a} = 1
\end{equation} 
which ensures that a vehicle cannot be assigned on two links in a certain time interval $t$.

\subsection{Solutions}
Considering the temporal continuous traffic flow, in time interval $t$, take the initial state $f^0_{k_a}$ in (\ref{equ:cong}) as $f^{t-1}_{k_a}$. In this paper, the test bench is based on Sioux Falls network, and the departing time is the same during each time interval $t$. Then, in time interval $t$, the DUE problem can be transfered as a SUE problem. The SUE is based on (\ref{equ: DUE1}), (\ref{equ:DUE2}), and (\ref{equ:DUE3}), and the Lagrangian can be generated as (\ref{eq:big_lagrangian_UTS}).    
\begin{figure*}[!t]
	%\scriptsize
	%\setcounter{equation}{15}
	\begin{equation}\label{eq:big_lagrangian_UTS}
	%\begin{split}
	 L_{1}(\theta,\lambda_1,\mu_1) = \sum_{k_a\in K_A}\sum_{t\in D_t} \int\limits^{\theta^t_{k_a}}_{0}f^t_{k_a}(w)dw +
	 \lambda^t_{1,k_a}\bigg[\theta^t_{k_a} - \sum_{ P_{r_us_u}}\sum_{d_1\in D_t} h^{d_1}_{p_{r_us_u}} \delta^{d_1t}_{p_{r_us_uk_a}}\bigg] + \mu^d_{1,r_us_u} \bigg[q^{d_1}_{r_us_u} - \sum_{ P_{r_us_u}} h^{d_1}_{p_{r_us_u}}\bigg]
	%\end{split}
	\end{equation}
	%\hrulefill
	\vspace*{0.0in}
\end{figure*}

The Hessian matrix of (\ref{equ: DUE1}) is positive definite, which indicate the convexity. The optimal travel flow $\theta^{*t}_{k_a}$ can be computed with the KKT conditions~\cite{boyd2004conv456ex, sheffy1985ur123ban}. Therefore, the solution of the UTS in time interval $t$ can be designed as in \textbf{Algorithm 1}.
\begin{algorithm}\label{algorithm1}
	\caption{DUE solution}
	\label{alg11}
	\begin{algorithmic}
		\STATE $\mathbf{Step\ 1}$: Initialization. Set the iteration number $n_{1} = 1$, the threshold $\varepsilon_{1}$. Update the new demand as $q^{dt}_{r_us_u}$ = $q^{dt-1}_{r_us_u}$. Then using \textit{all-or-nothing} assignment~\cite{sheffy1985ur123ban,chekuri200412all} to generate $\theta^t_{k_a}$.
		\STATE
		\STATE $\mathbf{Step\ 2}$: Update the travel time with $\theta^t_{k_a}$: $f^{t, n_{1}}_{k_a}$ = $f^{t}_{k_a}(\theta^{t, n_{1}}_{k_a})$, which is the beginning of the loop (Step 2 to Step 6).
		\STATE 
		\STATE $\mathbf{Step\ 3}$: Determine the descend direction. Find the shortest path with $f^{t, n_{1}}_{k_a}$ using \textit{all-or-nothing} assignment to generate $\phi^{t,n_{1}}$, which is the shortest route pattern.
		\STATE
		\STATE $\mathbf{Step\ 4}$: Determine the iteration step $\zeta^{n_1}$, which can be designed as:
		\begin{subequations}
			\begin{align}
			& min_{\zeta^{n_1}}\sum_{k_a}\int\limits_{0}^{\eta}f^{t}_{k_a}(\omega)d\omega \label{equ:step41} \\
			s.t. \ \ & \eta = \theta^{t, n_{1}}_{k_a} + \zeta^{n_1} (\phi^{t,n_{1}} - \theta^{t, n_{1}}_{k_a})  \label{equ:step42} \\
			& 0\leq \zeta^{n_1} \leq 1 \label{equ:step43}
			\end{align}
		\end{subequations}
	    The Golden-section search is employed to determine $\zeta^{n_1}$ in a short time.  
		\STATE
		\STATE $\mathbf{Step\ 5}$: Update $\theta^{t, n_{1}+1}_{k_a}$ with
		\begin{equation}
		\theta^{t, n_{1}+1}_{k_a} = \theta^{t, n_{1}}_{k_a} + \zeta^{n_1} (\phi^{t,n_{1}} - \theta^{t, n_{1}}_{k_a})
		\end{equation}
		\STATE
		\STATE $\mathbf{Step\ 6}$: Compare the result to the threshold $\varepsilon_{1}$ as:
		\begin{equation}
		\frac{\sqrt{\sum_{k_a}(\theta^{t, n_{1}+1}_{k_a} - \theta^{t, n_{1}}_{k_a})^2}}{\sum_{k_a}\theta^{t, n_{1}}_{k_a}} \leq \varepsilon_{1} \label{equ:DUE_thres}
		\end{equation}
		where if (\ref{equ:DUE_thres}) is fulfilled, the result of traffic flow in time interval $t$ is $\theta^{t, n_{1}+1}_{k_a}$; otherwise, go back to step 2, with $n_{1} = n_{1} + 1$.
	\end{algorithmic}
\end{algorithm}

\section{Optimal Power Flow of Power Distribution System (Lower Level)}\label{sec:PDS}
\subsection{Branch Flow Model}\label{sec:BFM}

The PDS can be represented in a radial graph as: $\mathcal{G}^{PDS} $= $[\mathcal{V}^{PDS}, \mathcal{E}^{PDS}]$, where $\mathcal{V}^{PDS}$ = $\{1,2,\cdots, n_{\mathcal{V}^{PDS}}\}$, and $\mathcal{E}^{PDS}$ = $\{1,2,\cdots, m_{\mathcal{E}^{PDS}}\}$. Similar as UTS, the PDS also contains the temporal characteristics, the superscript $(\cdot)^t$, $t\in D_t$. Considering the three-phase unbalanced system, the phase configuration is defined as $\phi $ $\subseteq$ $\{a,b,c\}$. For a radial distribution system, there is an unique $i^1$ as the ancestor bus for bus $i$, and the branch between bus $i^1$ and $i$ can be named as $i$~\cite{pe015dis456ed,low201454convex}. The branch flow model in time interval $t$ can be defined as:

\begin{subequations}\label{equ:constraint_PDS_BFM}
	\begin{align} 
	s^{\phi t}_i &= \text{diag}(S^{\phi t}_i - \sum_{j \in C_i}(S^{\phi t}_j-z^{\phi t}_jl^{\phi t}_j)),  \label{equ:eps1} \\
	v^{\phi t}_{i^1} &= v^{\phi t}_i-2(r^{\phi }_{i}P^{\phi t}_{i} + x^{\phi }_{i}Q^{\phi t}_{i}) + |z^{\phi }_{i}|^2l^{\phi t}_{i}, \label{equ:eps2}
	\end{align}
\end{subequations}
where $s^{\phi t}_i$ = $p^{\phi t}_{i}$ + $\textbf{i}q^{\phi t}_{i}$ indicates the power injection on bus $i$, $S^{\phi t}_i$ = $V^{\phi t}_i (I^{\phi t}_i)^H$ is the complex power from bus $i$ to bus $i^1$, $v^{\phi t}_{i}$ = $V^{\phi t}_{i} (V^{\phi t}_{i})^H$, $l^{\phi t}_{i}$ = $I^{\phi t}_{i} (I^{\phi t}_{i})^H$, $z^{\phi }_{i}$ =  $r^{\phi }_{i}$ + $\textbf{i} x^{\phi }_{i}$, $C_i$ is the set of buses that have a common ancestor bus $i$, and $H$ indicates the Hermitian transpose.

In addition, according to~\cite{gan2014con123vex}, the matrix $\mathbf{M}^t$ is defined as

\begin{equation}
\mathbf{M}^t = 
\begin{bmatrix}
v^{\phi t}_{i} & S^{\phi t}_i \\[0.3em] \label{equ:constraint_matrx}
(S^{\phi t}_i)^H & l^{\phi t}_{i}         
\end{bmatrix}
\end{equation}
where in (\ref{equ:constraint_matrx}), $\mathbf{M}^t\succeq0$ ( $\mathbf{M}^t$ is positive semidefinite) and $\mathbf{Rank}({\mathbf{M}^t})$ = 1.

\subsection{Objective Function and SDP Relaxation of OPF in PDS}\label{sec:OPF}
The objective function contains two major parts: the generation cost and system line loss, which can be designed as follows:
\begin{equation}
F^t_{PDS} = \sum_{i\in\mathcal{V^{PDS}}}\sum_{\phi_i}[\alpha_{1} (p^{\phi t}_{i})^2 + \beta_{1}(p^{\phi t}_{i})] \label{equ:constraint_PDS_obj}
\end{equation}
where $\alpha_{1}$ and $\beta_{1}$ are two weight coefficients for the generation cost and system line loss, respectively.

The constraints of the voltage $v^{\phi t}_{i}$ and current $l^{\phi t}_{i}$ are illustrated as follows:

\begin{subequations}\label{equ:constraint_PDS_v_l}
	\begin{align} 
	\underline{v^{\phi }_{i}} \leq v^{\phi t}_{i} \leq \overline{v^{\phi }_{i}},  \label{equ:constraint_PDS_v} \\
	\underline{l^{\phi }_{i}} \leq l^{\phi t}_{i} \leq \overline{l^{\phi }_{i}}, \label{equ:constraint_PDS_l}
	\end{align}
\end{subequations}
%where $\underline{v^{\phi }_{i}}$ and $\overline{v^{\phi }_{i}}$ are the lower and upper bounds of $v^{\phi }_{i}$, $\underline{l^{\phi }_{i}}$ and $\overline{l^{\phi }_{i}}$ are the lower and upper bounds of $l^{\phi }_{i}$.

In this paper, the injection power $s^{\phi t}_{i}$ plays an important role in active control and reactive control for the OPF.
\begin{equation}
\underline{s^{\phi }_{i}} \leq s^{\phi t}_{i} \leq P^{*t}_{i}, \label{equ:constain_PDS_s}
\end{equation}
where $P^{*t}_{i}$ is the optimal power injection in (\ref{equ:powerup}). Considering the relationship with the UTS, the optimal injection power $s^{*t}_{i_1}$ decides the number of EVs in CDS $i_1$ (Here for CDS, $i_1 = i$ for the PDS and UTS, which can be explained as Fig.~\ref{fig:testbench} with yellow circles). 

According to the SDP relaxation introduced in~\cite{gan2014con123vex,pe015dis456ed}, the constraint $\mathbf{Rank}({\mathbf{M}^t})$ = 1 of (\ref{equ:constraint_matrx}) can be removed, and the relaxed objective problem is shown as follows:

\begin{subequations}\label{equ:relax_obj_PDS}
	\begin{align} 
	&\min_{s^{\phi t}_{i}, v^{\phi t}_{i}, l^{\phi t}_{i}, S^{\phi t}_{i}} F^t_{PDS} \\
	s.t.\  &(\ref{equ:constraint_PDS_BFM}), (\ref{equ:constraint_PDS_v_l}), and\  \mathbf{M^t}\succeq0
	\end{align}
\end{subequations}
where (\ref{equ:relax_obj_PDS}) is a convex problem and can be solved with ADMM.

\subsection{OPF Solution with ADMM}
In the power-traffic system, the OPF computation of the three-phase unbalanced PDS is the bottleneck of the whole system operation. Firstly, because the real PDS usually contains a lot of feeders, which brings high computation loads for the system. Secondly, the PDS is a highly dynamic system, which requires a short computation time. Therefore, the ADMM, a distributed algorithm is proposed to solve this problem in a short time, which is based on Lagrange multiplier with an additional penalty term (augmentation term). 

The standard form of the ADMM can be formulated as follows:

\begin{subequations}\label{equ:relax_org_ADMM}
	\begin{align} 
	&\min_{x,y} f_{A}(x) + g_{A}(y) \\
	s.t. &\  x \in \mathcal{K}_x, y \in \mathcal{K}_y, and \ x = y
	\end{align}
\end{subequations}
where $f_{A}$ and $ g_{A}$ are convex, and $\mathcal{K}_x$ and $\mathcal{K}_y$ are convex sets. With an additional penalty term $\rho/2 ||x-y||^2_2$, the efficient augmented Lagrangian function is shown as follows~\cite{boyd2011d45istributed}:

\begin{equation}\label{equ:relax_aug_ADMM}
%\scriptsize 
%\begin{align}
L_{\rho}(x,y,\lambda) = f_{A}(x) + g_{A}(y) + <\lambda,x-y> + \frac{\rho}{2} ||x-y||^2_2,
%\end{split}
%\end{align}
\end{equation} 
where $0\leq \rho$ is a coefficient for convergence, and $\frac{\rho}{2} ||x-y||^2_2$ is the quadratic penalty term to increase the converge speed.

Therefore, the relaxed objective function (\ref{equ:relax_obj_PDS}) can be formulated as 

\begin{subequations}\label{equ:ADMM_advance}
	%\scriptsize
	\begin{align} 
	& \min_{S^t_i, s^t_i, v^t_i, l^t_i} F^t_{PDS} = \min_{S^t_i, s^t_i, v^t_i, l^t_i} \sum_{i \in \mathcal{V^{PDS}}} F^t_{PDS,i} \label{equ: ADMM_ad1}\\
	& s.t. \ (\ref{equ:constraint_PDS_BFM}), (\ref{equ:constraint_PDS_v_l}), and\  \mathbf{M}\succeq0 \label{equ: ADMM_ad2}\\
	& S^{\phi t,(x)}_{i,i^1} = S^{\phi t,(y)}_{i}, l^{\phi t,(x)}_{i,i^1} = l^{\phi t,(y)}_{i}, v^{\phi t,(x)}_{i,i^1} = v^{\phi t,(y)}_{i} \label{equ: ADMM_ad3}\\
	& S^{\phi t,(x)}_{i} = S^{\phi t,(y)}_{i},  l^{\phi t,(x)}_{i} = l^{\phi t,(y)}_{i},  v^{\phi t,(x)}_{i} = v^{\phi t,(y)}_{i} \label{equ: ADMM_ad4}\\ 
	& s^{\phi t,(x)}_{i} = s^{\phi t,(y)}_{i} \label{equ: ADMM_ad5}
	\end{align}
\end{subequations}

where the variables $(S^t_i, s^t_i, v^t_i, l^t_i)$ is maintained in bus $i$, which can be seemed as an local agent. In the ADMM, the $x$-update and $y$-update indicate as $( \cdot )^{(x)}$ and $( \cdot )^{(y)}$.

Then, as in~\cite{pe015dis456ed,boyd2011d45istributed} the augmented Lagrangian of (\ref{equ:ADMM_advance}) can be derived as (\ref{eq:big_lagrangian}).

\begin{figure*}[!t]
	%\scriptsize
	%\setcounter{equation}{15}
	\begin{subequations}\label{eq:big_lagrangian}
		\begin{align}
		L^t_{\rho}(x^t,y^t,\lambda^t)	
		& = \sum_{i \in \mathcal{V^{PDS}}}F^t_{PDS,i} + <\lambda^t, x^t_i-y^t_i> + <\mu^t_i, x^t_{i,i^1} - y^t_i> + \sum_{j \in C_i}<\gamma^t_j, v^{(x)t}_{i,j} - v^{(y)t}_i> + \frac{\rho}{2} f^t_{PDS,i,y}\\
		& = \sum_{i \in \mathcal{V^{PDS}}}F^t_{PDS,i} + <\lambda^t, x^t_i-y^t_i> + \sum_{j \in C_i} <\mu^t_j, x^t_{j,i} - y^t_j> + <\gamma^t_i, v^{(x)t}_{i^1,i} - v^{(y)t}_{i^1}> + \frac{\rho}{2} f^t_{PDS,i,x}
		\end{align}
	\end{subequations}
	%\hrulefill
	\vspace*{0.0in}
\end{figure*}
The detail formulation of $f^t_{PDS,i,y}$ and $f^t_{PDS,i,x}$ can be found in~\cite{pe015dis456ed}, and at iteration $k$, the variables $x$, $y$, and $\lambda$ are illustrated as follows:

\begin{subequations}\label{equ:ADMM_updates}
	\begin{align} 
	& x^{t,k+1} \in \arg\min_{x \in \mathcal{K}_x} L_{\rho}(x^t,y^{t,k},\lambda^{t,k}) \label{equ:ADMM_updates_x} \\
	& y^{t,k+1} \in \arg\min_{y \in \mathcal{K}_y} L_{\rho}(x^{t, k+1},y^t,\lambda^{t,k}) \label{equ:ADMM_updates_y} \\
	& \lambda^{t,k+1} = \arg\min_{x \in \mathcal{K}_x} \lambda^{t,k} +\rho(x^{t, k+1} - y^{t, k+1}) \label{equ:ADMM_updates_ru}
	\end{align}
\end{subequations}
Considering the time intervals of the UTS, the ADMM-based distributed OPF computation architecture is built to dramatically reduce the time consumption to the proposed power-traffic coordinated operation approach.

Therefore, in addition to the designed user equilibrium in the higher level, the detail models of UTS and PDS are build to provide the spatial and temporal information in the lower level. In the next section, a smart charging/discharging strategy is proposed for the EVs and bidirectional CDSs to further reduce the impact to PDS.

\section{Smart EV Charging/discharging (Lower Level)}\label{Sec:EVs}

In this paper, the EVs and CDSs act as the reserves for both PDS and UTS. In this section, we take CDS $i_1$ as an example, other CDSs contain the same performances. In time interval $t$, $n^t_{i_1}$ EVs are parked in CDS $i_1$ with parking time $\tau_2$, which is defined in (\ref{equ:ISO}). In real world, $n^t_{i_1}$ depends on many factors, for example, different habits of the drivers, which are beyond the scope of this paper. Here, $n^t_{i_1}$ is determined as follows: 
\begin{equation}\label{equ:park_EV}
%\scriptsize min{P^{*t}_{i_1}/C^t_{1}, \theta^t_{k_a}}  \varrho_3 * min{P^{*t}_{i_1}/C^t_{1}, \theta^t_{k_a}}
\begin{split}
& n^t_{i_1} =  \\
&\left\{ \begin{array}{ll}
\varrho_3 *\min\{s^{*t}_{i_1}/C^t_{1}, \theta^t_{k_a}\} & \mbox{If $\pi^{*t} \geq \varrho_1 + \varrho_2$} \\
\varrho_4 * \min\{s^{*t}_{i_1}/C^t_{1}, \theta^t_{k_a}\}  & \mbox{If $ \varrho_1 \leq \pi^{*t} \leq \varrho_1 + \varrho_2$}  \\
0    & \mbox{If $\varrho_1 \geq \pi^{*t}$}
\end{array}
\right.
\end{split}
\end{equation}
where $ s^{*t}_{i_1}$ is the optimal power injection, which is determined in the OPF solution (\ref{equ:ADMM_updates}). $C^t_{1}$ is the average discharging/charging speed of EVs. $\varrho_1$ is the electrical power price, $\varrho_2$ is the congestion fee, $\varrho_3$ is the ratio for EV parking to the CDSs. $n^t_{i_1} < \chi_{i_1}$, $\chi_{i_1}$ is the capacity of CDS $i_1$.
 
During the parking time $\tau_2$, the stochastic EVs charging/discharging can impact the stability of the PDS. Considering the SOC, a smart EVs charging/discharging approach is proposed to reduce the impact to PDS as follows:

\begin{subequations}\label{equ:smart_EV}
	%\scriptsize
	\begin{align}
	F^t_{CDS} = & \min_{C^{t_1}_{ev, i_3}} \sum_{t_1\in \tau_2} \bigg[P^{t_1+1}_{CDS} (C^{t_1}_{ev, i_3}) - P^{t_1}_{CDS} (C^{t_1}_{ev, i_3}) \bigg]^2  \label{equ: CDS1}\\
	s.t.\ \  & \underline{C_{ev}} \leq C^{t_1}_{ev, i_3} \leq \overline{C_{ev}} \label{equ: CDS2}\\
	& P^{t_1}_{CDS} (C^{t_1}_{ev, i_3}) = Q^{t_1}_{PDS} - \sum^{n^t_{i_1}}_{i_3} C^{t_1}_{ev, i_3} \label{equ: CDS3}\\
	& Q^{\tau_2}_{ev, i_3} = \sum_{t_1\in \tau_2}C^{t_1}_{ev, i_3}, \ 0 \leq i_3 \leq n^t_{i_1}  \label{equ: CDS4}\\
	&  \underline{\Upsilon} \leq SOC \leq  \overline{\Upsilon}, \  \label{equ: CDS5} 
	\end{align}
\end{subequations} 
where $C^{t_1}_{ev, i_3}$ is the discharging (positive) and charging (negative) speed of EV $i_3$ at time $t_1$. In (\ref{equ: CDS3}), $P^{t_1}_{CDS} (C^{t_1}_{ev, i_3})$ is the netload at CDS $i_1$, which equals to original load of PDS $Q^{t_1}_{PDS}$ minus the total charging/discharging power. In (\ref{equ: CDS4}), $Q^{\tau_2}_{ev, i_3}$ is the total demand of charging/discharging of EV $i_3$, which contains a SOC constraint as (\ref{equ: CDS5}),  $\Upsilon$ is the ratio of SOC. 
%Specifically, it is assumed that all the EV has the similar capacity and similar charging/discharging speed, the average charging/discharging speed of CDS $i_1$ can be computed as (\ref{equ: CDS6}).

\section{Numerical Simulation and Results}
\label{sec:Implementation}
\begin{figure*}[t!]
	\begin{center}
		\includegraphics [width=1.00\columnwidth, angle=90]{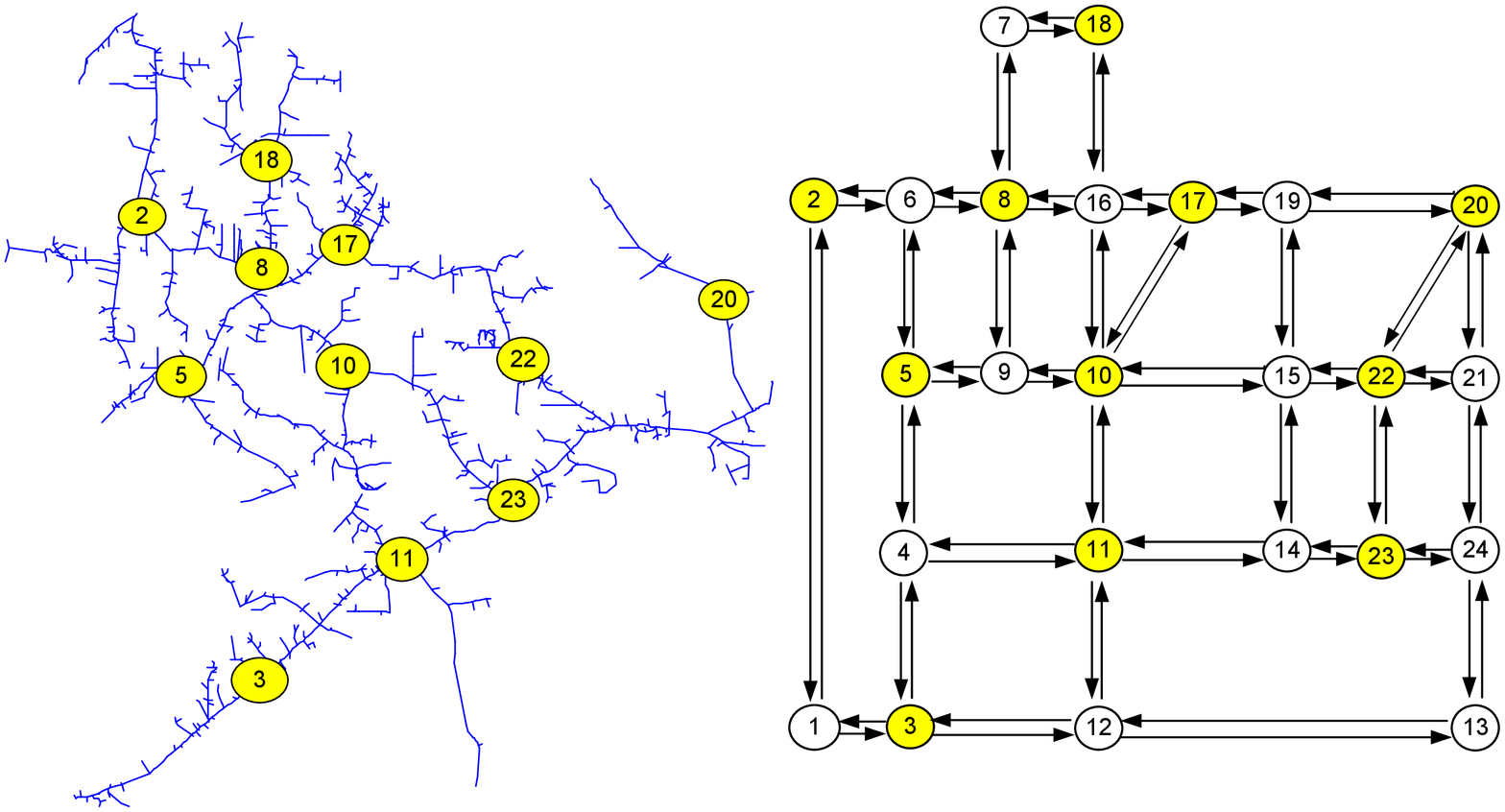}
		\caption{The test bench consists of a PDS and a UTS.}\label{fig:testbench}
	\end{center}
\end{figure*}
As shown in Fig.~\ref{fig:testbench}, the test bench is based on the IEEE 8,500-bus PDS and the Sioux Falls UTS~\cite{IEEE8500PDS,lam2008model123ing}, which cover the similar geographical areas. It is assumed that a small city with 10,000 EVs has the bi-peak and bi-ramp problem, the "duck curve" data is based on~\cite{Duck2017c} and the traffic behavior data is based on~\cite{gonzal65es2013evening}. 11 CDSs are located in node 2, 3, 5, 8, 10, 11, 17, 18, 20, 22, 23, which are shown with yellow circles and illustrated in both PDS and UTS, respectively. The time interval $t$ is 15 minutes for the power-traffic systems. The PDS cost function $f^t_{PDS}$ is defined as quadratic function as in~\cite{park2005particle}, and the UTS cost function $f^t_{UTS}$ is defined similarly. Both $f^t_{PDS}$ and $f^t_{UTS}$ are convex, and the quadratic form indicates the utility cost increase fast with the increasing of PDS load and UTS load. As discussed above, the utility objective function $F^t_{UTY}$ is convex.  The weight factor $\gamma_1$ equals to 1, which indicates the same importance of PDS and UTS. Similarly, the customer objective function $F^t_{CSO}$ is also set as a convex function. The electrical power price $\varrho_1$ = $45 \ \$/MWh $, and the congestion fee $\varrho_2$ = $2 \ \$/h$ \cite{EIA2017132US,de2011traf78fic}. The EVs parking ratio $\varrho_3$ = $0.8$, $\varrho_4$ = $0.3$. The parking capacity for each CDS is $\chi_1$ = 100. The simulations are executed using a server with 3.60 GHz Intel Xeon CPU and 32 GB RAM, and the software are Python, MATLAB, MATLAB global optimization toolbox, and parallel computing toolbox. The communication network is good enough to transmit all the information in real-time without any cybersecurity issues. 

\subsection{Flowchart of the Simulation}
\begin{figure}[h!]
	\begin{center}
		\includegraphics [width=0.80\columnwidth]{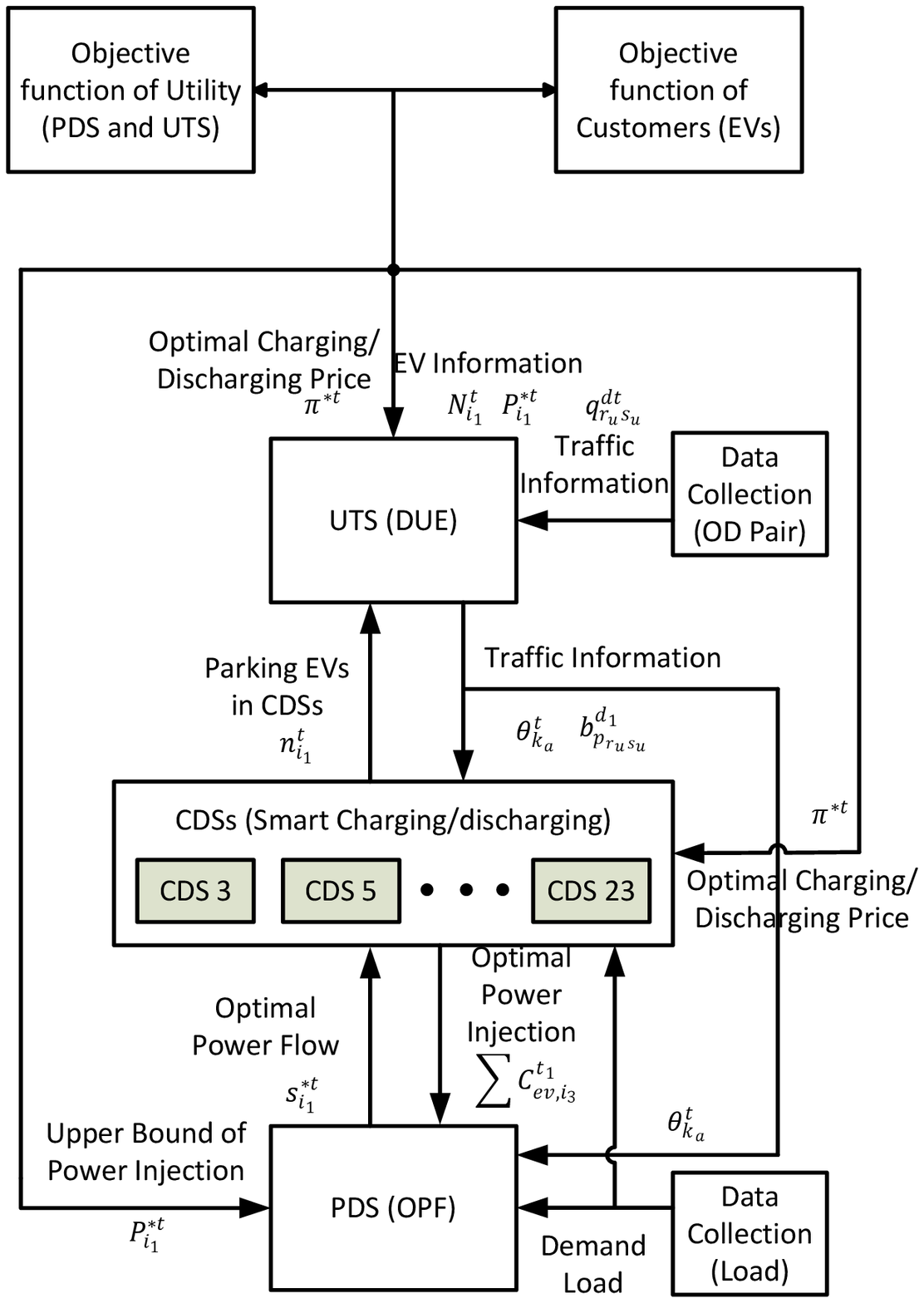}
		\caption{The test bench consists of a PDS and a UTS.}\label{fig:simuflow1}
	\end{center}
\end{figure}

\begin{figure}[t!]
	\begin{center}
		\includegraphics [width=0.8\columnwidth]{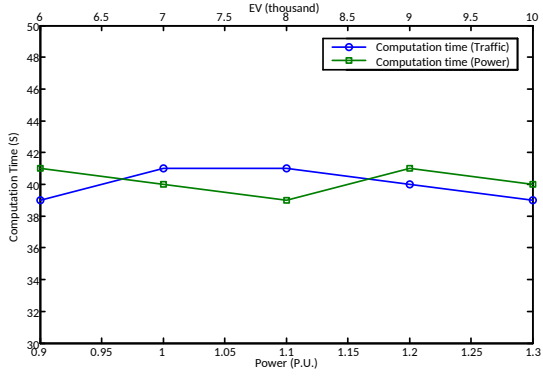}
		\caption{ Computation time comparison with different traffic and load factor.}\label{fig:Duck_curve2}
	\end{center}
\end{figure}

The simulation flowchart is shown in Fig.~\ref{fig:simuflow1}, and the corresponding description is shown in \textbf{Algorithm 2}. In Step 2, a distributed algorithm is used to reduce the time consumption of the equilibrium computation. In Step 3, for each time interval $t$, the DUE problem is transferred as a SUE problem. The the Golden-section search is employed to determine the iteration step $\zeta^{n_1}$ in a short time for the UTS with about 10,000 EVs and 11 CDSs. In Step 4, because the test bench is based on the IEEE 8,500-bus PDS, the ADMM is implemented to computed the OPF in distributed manner with the constraints from Step 3. In Step 5, for each CDS, the charging/discharging are independent, the smart EVs charging/discharging can be computed in parallel, which also helps to reduce the computation time. The computation time analysis is shown in Fig~\ref{fig:Duck_curve2}. The two curves indicate the average computation times of the proposed approach are less than 50 s with 100 scenarios. Compared with the time interval 15 minutes, the proposed approach is quick enough to support continuously operations for the power-traffic systems.
\begin{algorithm}\label{algorithm2}
	\caption{Simulation Process}
	\label{alg2}
	\begin{algorithmic}
		\STATE $\mathbf{Step\ 1}$: Initialization and data collection. Collect the data and parameters of the PDS, UTS, CDSs, and EVs, such as the topology information, electrical power price $\varrho_1$, congestion fee $\varrho_2$, set the simulation time interval $t$ = 15 min, etc.
		\STATE
		\STATE $\mathbf{Step\ 2}$: Higher level: equilibrium computation for the defined utility part (PDS \& UTS) and customer part (CDSs \& EVs). Build the utility and customer objective function as (\ref{equ:ISO}) and (\ref{equ:obj_cus}). Then, compute the equilibrium $\pi^{*t}$ and $P^{*t}_{i_1}$ with (\ref{equ:opti}) and (\ref{equ:equ_iter}).
		\STATE 
		\STATE $\mathbf{Step\ 3}$: Lower level: DUE in UTS. Receive the information such as $\pi^{*t}$, $N^t_{i_1}$, $q^{d_1t}_{r_us_u}$ and $P^{*t}_{i_1}$, build the DUE objective function with constraints as in (\ref{equ:due_equa}), solve it with \textbf{Algorithm 1}, and generate the traffic information such as $\theta^t_{k_a}$ and $b^{d_1}_{r_us_u}$.
		\STATE
		\STATE $\mathbf{Step\ 4}$: Lower level: OPF in PDS. Receive the information such as $\pi^{*t}$, $\theta^t_{k_a}$, and $P^{*t}_{i_1}$, build the OPF objective function as in (\ref{equ:constraint_PDS_obj}) with the constraints (\ref{equ:constraint_PDS_v_l}) and (\ref{equ:constain_PDS_s}), relax it as (\ref{equ:relax_obj_PDS}), solve it as (\ref{eq:big_lagrangian}) and (\ref{equ:ADMM_updates}). Then, generate the $s^{*t}_{i_1}$ to CDS.
		\STATE
		\STATE $\mathbf{Step\ 5}$: Lower level: smart EVs charging/discharging. Receive the information such as $s^{*t}_{i_1}$, $\pi^{*t}$, and $N^t_{i_1}$, compute the parking EVs for each CDS as (\ref{equ:park_EV}), build the smart charging/discharging as (\ref{equ:smart_EV}), then generate the feedback information $n^t_{i_1}$ for UTS and $\sum^{n^t_{i_1}}_{i_3} C^{t_1}_{ev, i_3}$ for PDS. 
		\STATE
		\STATE $\mathbf{Step\ 6}$: The power-traffic system update with the new UTS and PDS information, then back to Step 2 with $t$ = $t+1$.
	\end{algorithmic}
\end{algorithm}

\subsection{Bi-Peak Shaving and Bi-Ramp Smoothing}
\begin{figure}[!t]
	\begin{center}
		\subfigure[]{ \label{fig:Duck_curve3}
			\resizebox{2.8in}{!}{\includegraphics{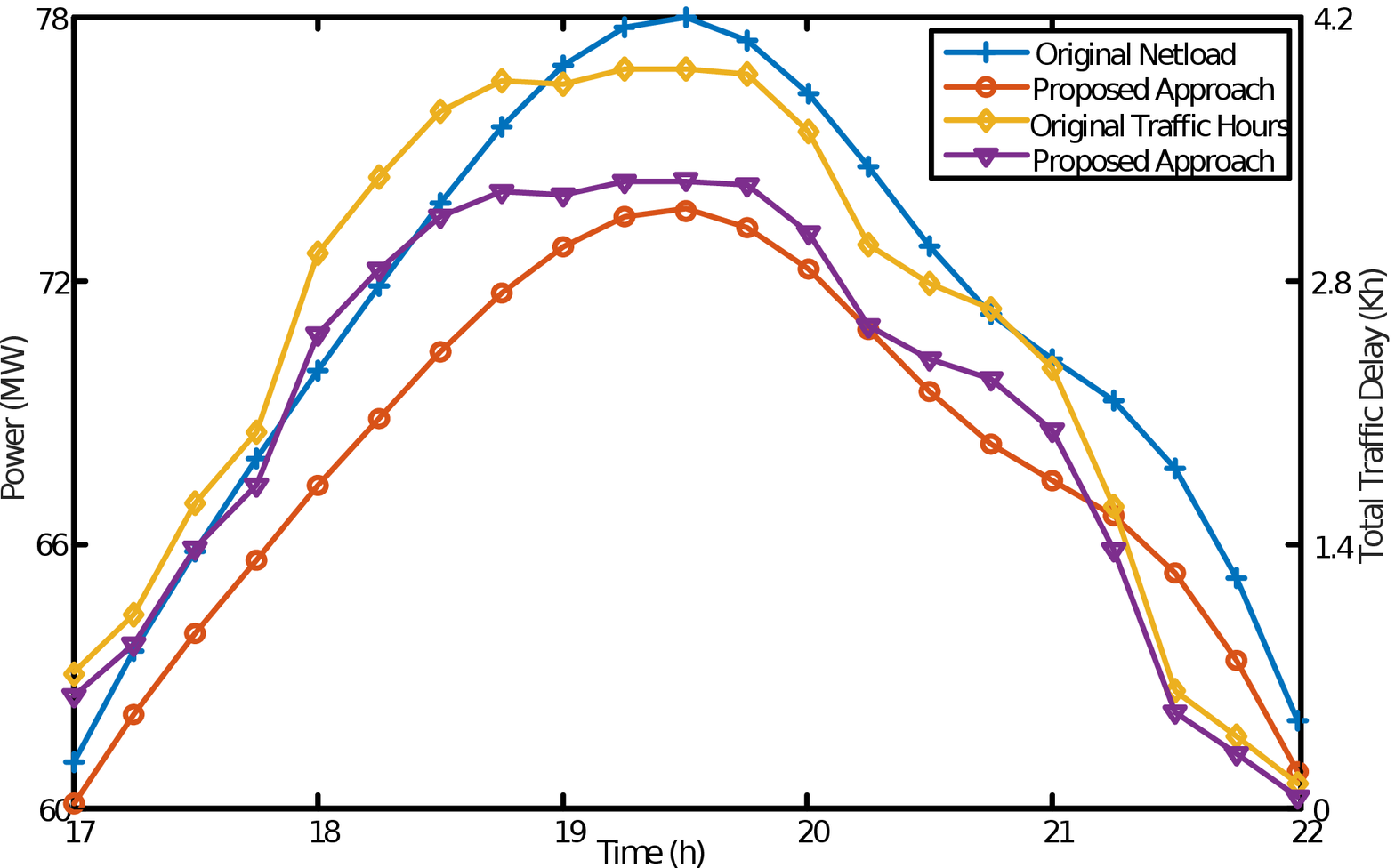}}}
		\subfigure[]{ \label{fig:Duck_curve4}
			\resizebox{2.8in}{!}{\includegraphics{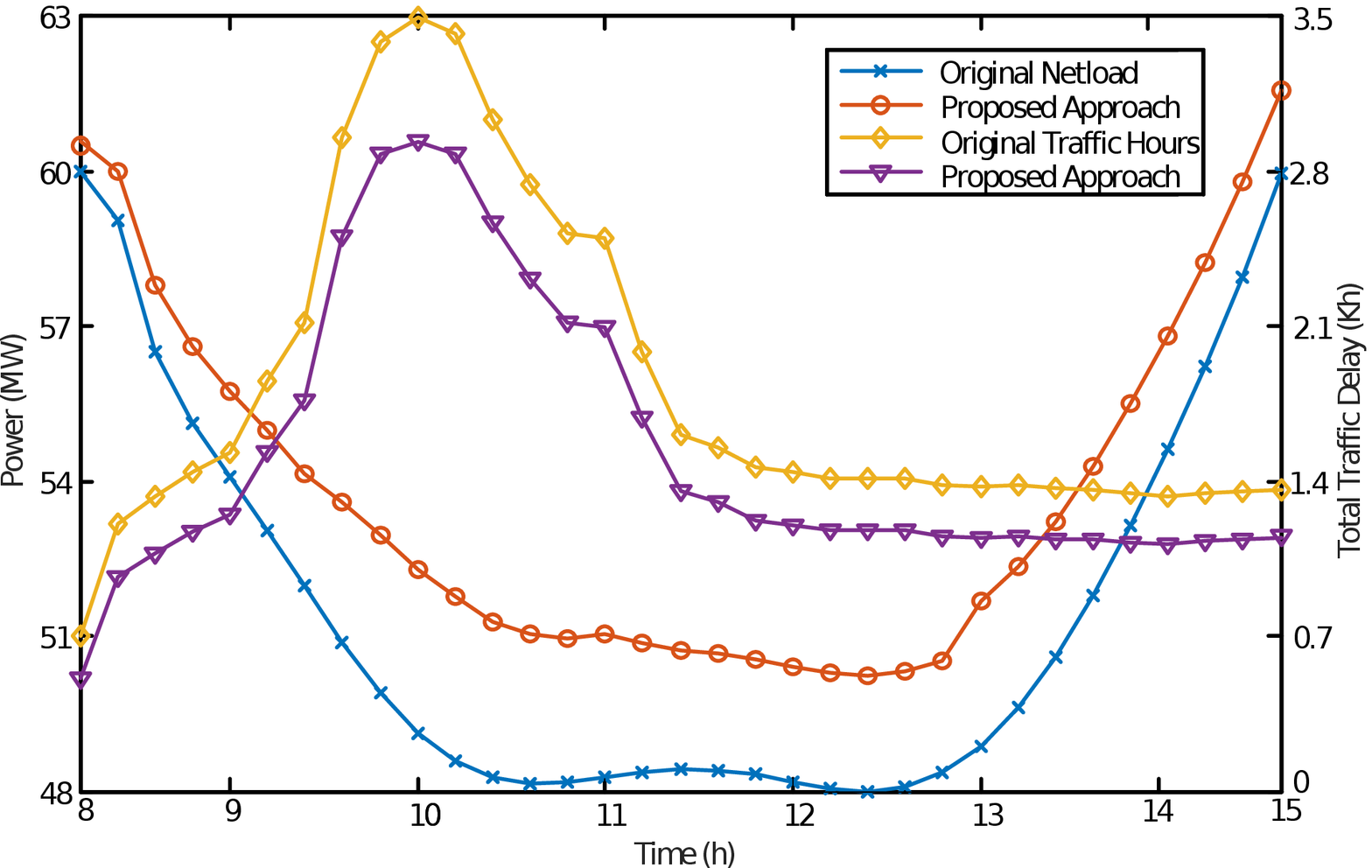}}}
		\caption{(a) The bi-peak shaving and bi-ramp smoothing for PDS and UTS from 17:00 to 22:00, (b) The peak-shaving and ramp-smoothing for UTS and over-generation compensation for PDS from 8:00 to 15:00.}\label{fig:Duck_curve5}
	\end{center}
\end{figure}

As shown in Fig.~\ref{fig:Duck_curve3}, the blue curve is the netload peak of PDS from 17:00 to 22:00, and the orange curve is the result of proposed approach. The peak load decreases from 78 MW to 72 MW with 15\% EVs discharging. The yellow curve is the traffic delay peak in UTS, and the purple curve is the reduced traffic delay with the proposed approach. The total peak traffic delay time of UTS decreases from about 4,100 hours to 2,900 hours. In Fig.~\ref{fig:Duck_curve4}, it is also clear that from 8:00 to 15:00, the netload of PDS increases from 48 MW to 51 MW, and the total traffic delay of UTS decreases from 3,500 hours to 2,800 hours. From Fig.~\ref{fig:Duck_curve5}, it is clear that the proposed approach can benefit both PDS and UTS to reduce the bi-peak and smooth the bi-ramp.

\begin{figure}[!t]
	\begin{center}
		\subfigure[]{ \label{fig:Duck_curve9}
			\resizebox{2.8in}{!}{\includegraphics{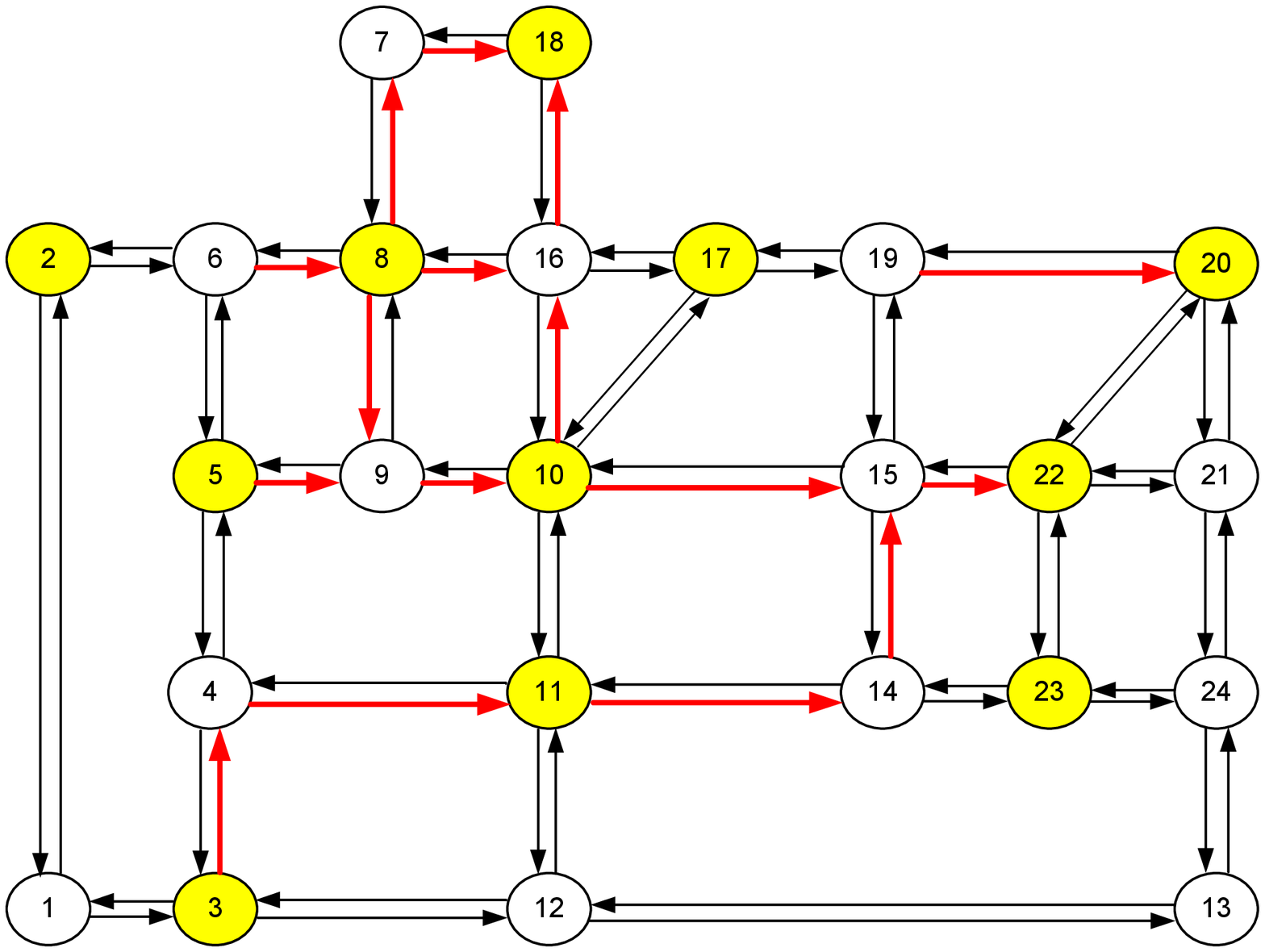}}}
		\subfigure[]{ \label{fig:Duck_curve10}
			\resizebox{2.8in}{!}{\includegraphics{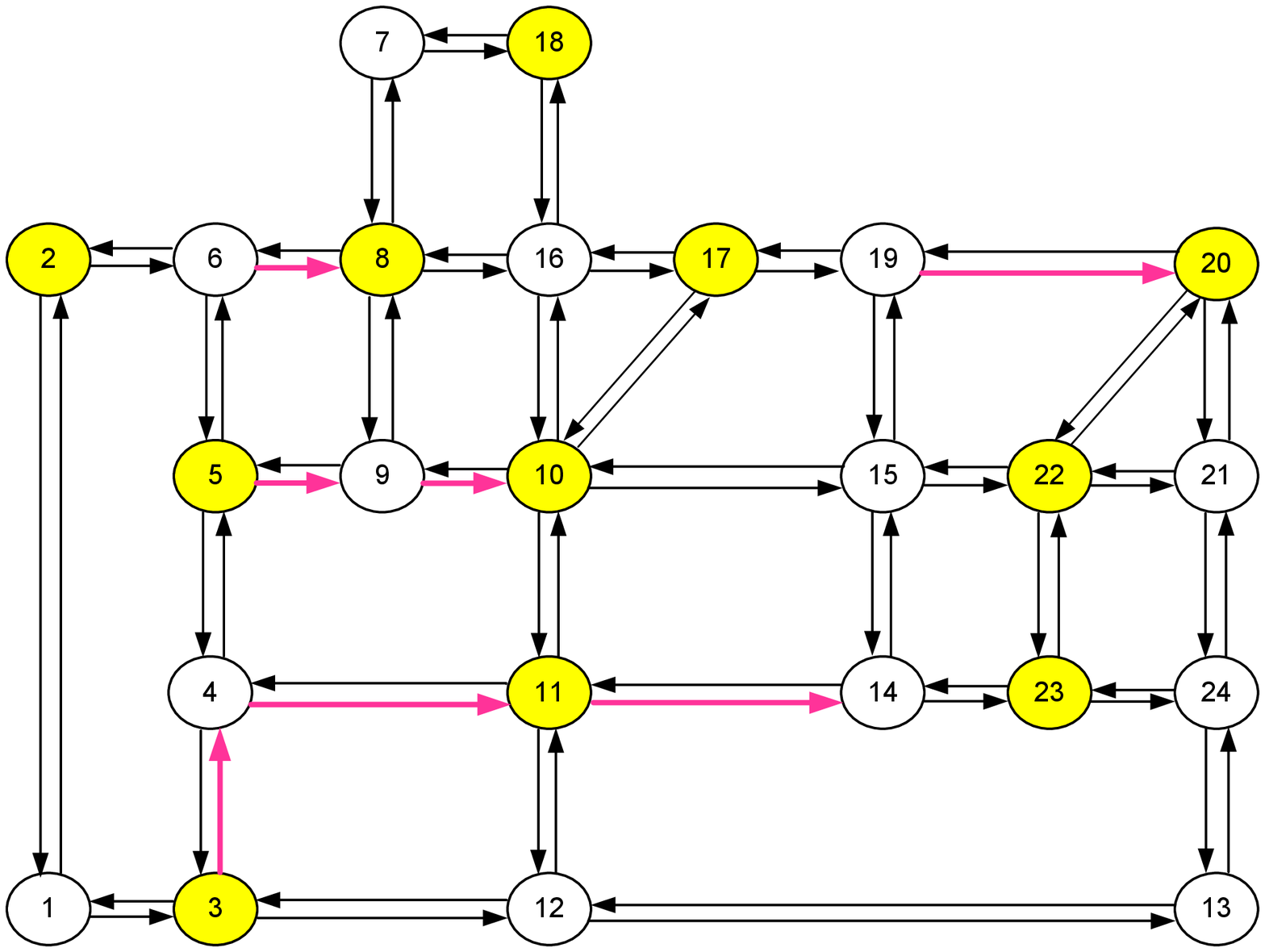}}}
		\caption{(a) The traffic congestion scenario in 18:30, (b) The traffic congestion scenario with proposed approach in 18:30.}\label{fig:Duck_curve8}
	\end{center}
\end{figure}

In Fig.~\ref{fig:Duck_curve8}, the detailed traffic congestion information of a congestion scenario is shown with the traffic maps. The OD pair of this scenario is shown in Table~\ref{Table: OD pair}. The red and pink arrows indicate the traffic congestions in the roads. The traffic congestion information without the proposed approach is shown in Fig.~\ref{fig:Duck_curve9} in 18:30. The red arrows indicate that the traffic delays on these roads are larger than 15 minutes. As shown in Fig.~\ref{fig:Duck_curve10}, with the proposed approach, the pink arrows indicate that the traffic delays on these roads are less than 15 minutes. The number of pink arrows (6 pink arrows) is much less than the number of red arrows (16 red arrows), which also indicates the alleviation of traffic congestion with the proposed approach. The total congestion delay reduces from 4100 hours to 2900 hours.

In summary, from Fig.~\ref{fig:Duck_curve5} and Fig.~\ref{fig:Duck_curve8}, the bi-peak and bi-ramp problems can be reduced and smoothed by the proposed approach, which benefits both PDS and UTS.
\begin{center}
	\begin{table}[h]
		%\scriptsize
		\caption{The OD pair of UTS}\label{Table: OD pair}\vspace{-0.1in}
		\begin{center}
			\begin{tabular}{ | c | c | c | c | c | c | c |}
				
				\hline
				Nodes & 13  & 14  &  18 &  20 &  21 & 23  \\ \hline
				1 	  & 200 & 350 & 240 & 300 & 256 & 345 \\ \hline
				2 	  & 270 & 210 & 360 & 200 & 200 & 310 \\ \hline
				3 	  & 200 & 300 & 220 & 345 & 270 & 345 \\ \hline
				5 	  & 250 & 320 & 260 & 450 & 230 & 345 \\ \hline
				6 	  & 300 & 210 & 270 & 250 & 300 & 300 \\ \hline
				10 	  & 200 & 200 & 345 & 345 & 200 & 300 \\ \hline
			\end{tabular}
		\end{center}
	\end{table}
\end{center}

\subsection{Social Cost Analysis}
\begin{figure}[t!]
	\begin{center}
		\includegraphics [width=0.80\columnwidth]{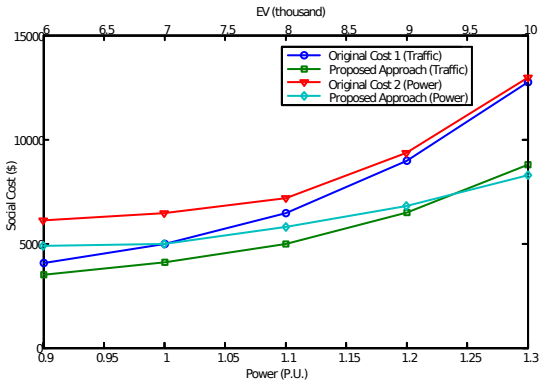}
		\caption{The different social costs comparison with different traffic and load factors.}\label{fig:Duck_curve1}
	\end{center}
\end{figure}

In Fig.~\ref{fig:Duck_curve1}, the blue curve with circle and green curve with square are in one group, which indicates the social costs with different number of EVs on the road. The load of PDS is 1.0 P.U., which means the total load of PDS is 60 MW. From Fig.~\ref{fig:Duck_curve1}, the cost of the proposed approach is lower, and it is clear that the proposed approach can benefit more for the social cost with the increasing number of EVs. In Fig.~\ref{fig:Duck_curve1}, the red curve and cyan curve indicate the social cost with different power factors. The EV number is 8000 in the UTS. From Fig.~\ref{fig:Duck_curve1}, it is also clear that the proposed approach can benefit more for the social cost with the increasing power factor. In summary, the prosed approach can benefit both PDS and UTS with different traffic scenarios and load factors. In addition, the social cost increases more with the increasing of PDS load or EV number, which indicates the quadratic objective function of PDS and UTS. This design benefits the power-traffic system to reduce the total social cost. 

%\begin{figure}[!t]
%	\begin{center}
%%		\subfigure[]{ \label{fig:Duck_curve1}
%%			\resizebox{2.6in}{!}{\includegraphics{figure/Social_cost.eps}}}
%		%
%		\subfigure[]{ \label{fig:Duck_curve2}
%			\resizebox{2.6in}{!}{\includegraphics{figure/Computation_time.eps}}}
%		%
%		\caption{(a) The different social cost comparison with different traffic and load factor, (b) Computation time comparison with different traffic and load factor.}\label{fig:Duck_curve}
%	\end{center}
%\end{figure}

\subsection{Smart EVs Charging/Discharging} 
\begin{figure}[!t]
	\begin{center}
		\subfigure[]{ \label{fig:smart_EVs1}
			\resizebox{2.8in}{!}{\includegraphics{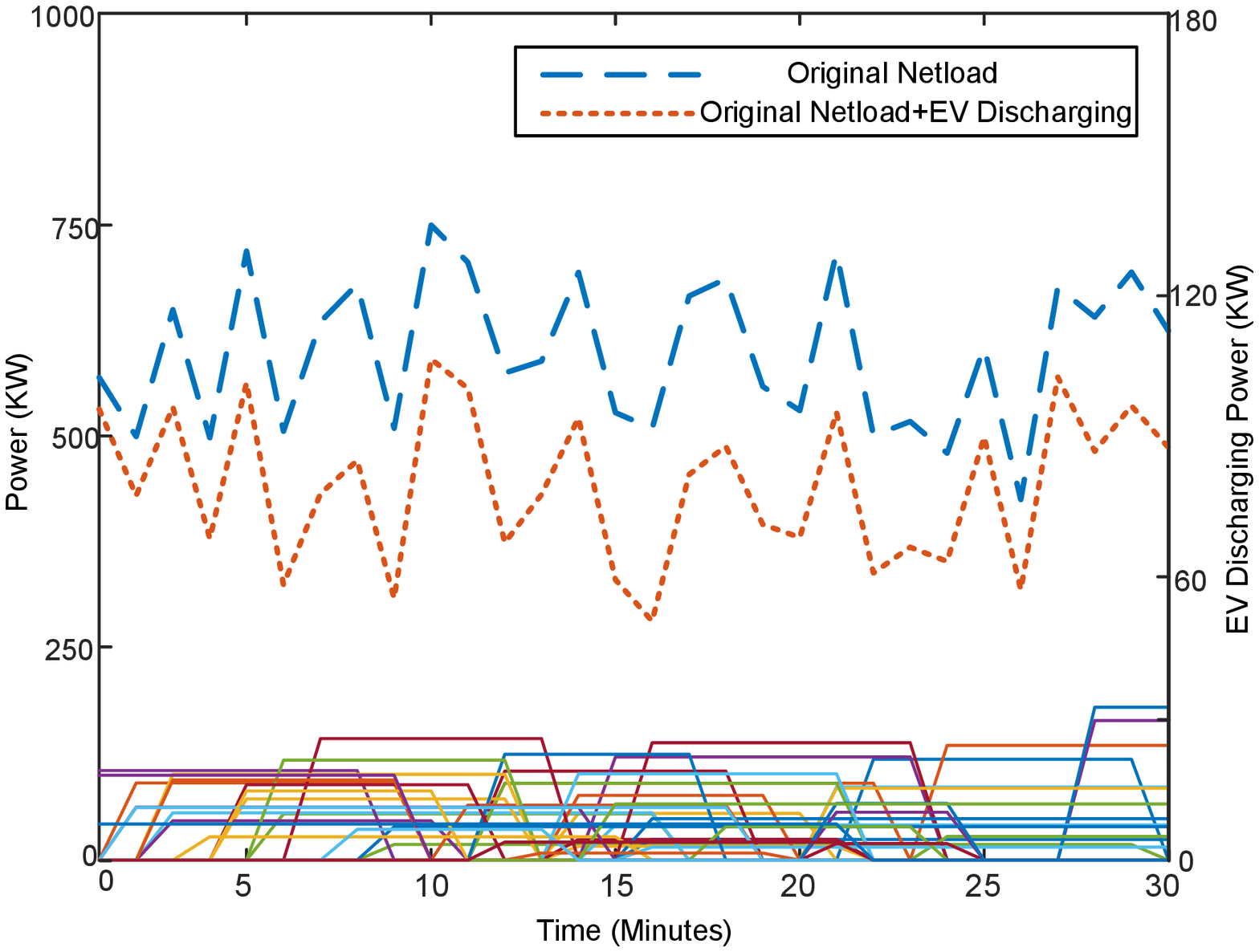}}}
		\subfigure[]{ \label{fig:smart_EVs2}
			\resizebox{2.8in}{!}{\includegraphics{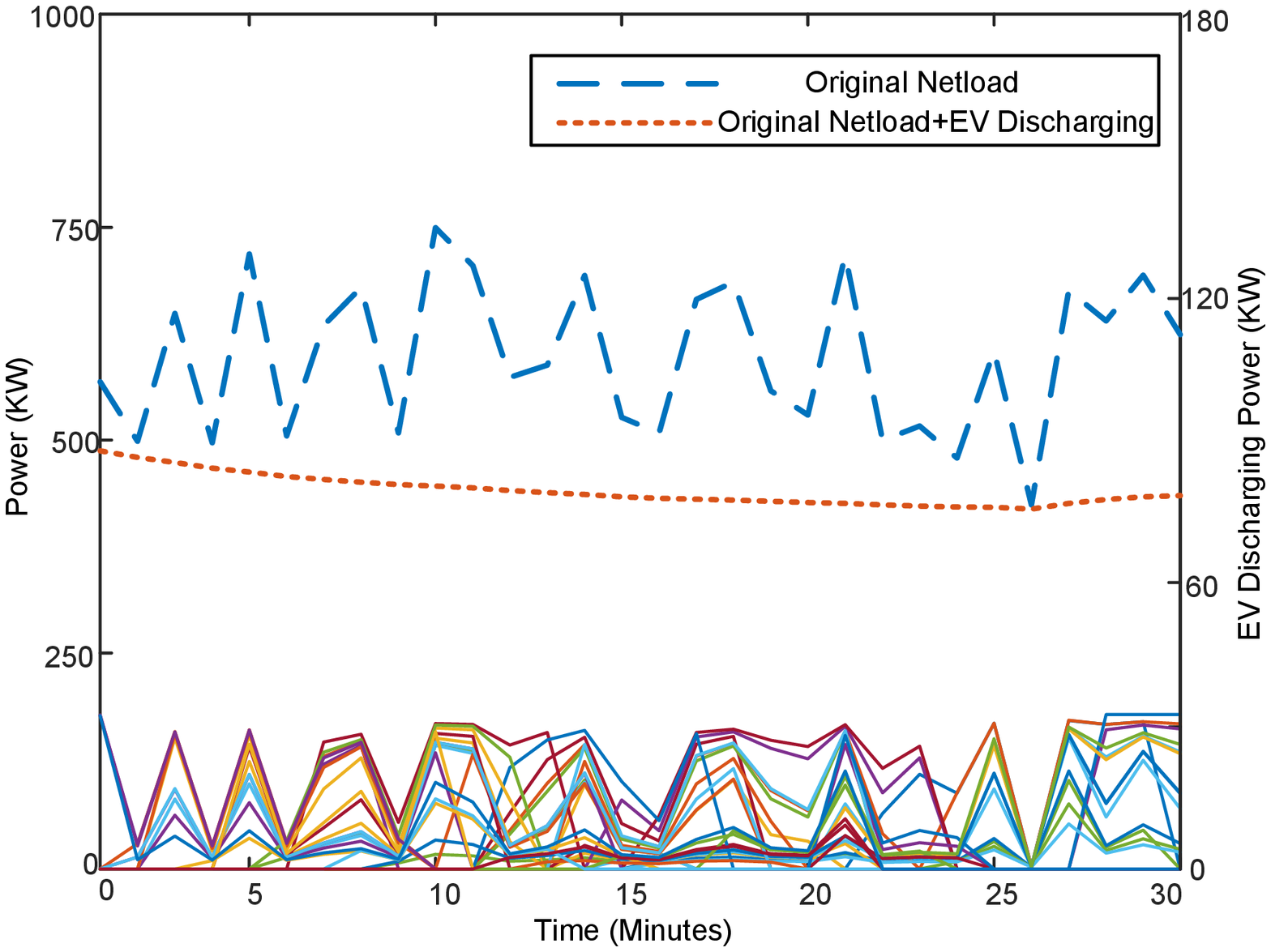}}}
		\caption{(a) The normal EV discharging in CDS 3 without considering the impacts to the PDS, (b) smart EV discharging in CDS 3 considering the impacts to the PDS.}\label{fig:smart_EVs}
	\end{center}
\end{figure}
To test the robustness of the proposed approach, we select two continuous intervals load data, which contains the biggest deviations within the 30 minutes. In Fig.~\ref{fig:smart_EVs}, EV discharging is taken as an example and 50 EVs are employed to test the proposed approach. As shown in Fig.~\ref{fig:smart_EVs1}, without the proposed approach, the stochastic EV discharging behaviors increase the deviations of the netload. In Fig.~\ref{fig:smart_EVs2}, with the proposed smart charging/discharging approach, the deviations of the netload are smoothed, which decreases the impacts to the PDS during the EV discharging. At the same time, the discharging speeds are controlled to meet the constraint (\ref{equ: CDS2}).

\section{Conclusion}\label{Sec:Concl}
In this paper, a hierarchical approach is proposed to shave the bi-peak and smooth the bi-ramp problem in the power-traffic system from higher level to lower level. The higher level gives an ``overview" for the whole system, and the lower level gives the detailed description for each part. The EVs and CDSs function as reserves for both the PDS and UTS to utilize the flexibility and optimize the operations of the power-traffic system.  In the higher level, the PDS and UTS are treated together to minimize the social cost, and the EVs and CDSs are treated as customers to minimize their expenditure. Then, an equilibrium is designed to determine the optimal charging/discharging prices, total demand electrical power, and total required EVs. In the lower level, considering the spatial and time domain, the detail models of PDS and UTS are built to specifically determine the power injection and EVs behaviors. In each CDS, a smart EVs charging/discharging approach is proposed to further reduce the impacts to the PDS. In the numerical results, the test bench consists of the IEEE 8,500-bus PDS and Sioux Falls UTS with about 10,000 EVs and 11 CDSs, which are used to demonstrate the feasibility and effectiveness of the proposed approach.

In real-world implementation, the weather, human behavior, and social issues bring the stochastic impacts to the power-traffic system, which increase the uncertainties and bring more challenges for system operation. In addition, the other systems such as the natural gas delivery system and the water system can also impact the proposed system and result in nonnegligible consequences. In the next step, other factors such as stochastic of renewable energies, the departure time of EVs, cybersecurity, multi-energy system, and human behaviors will be taken into consideration.

\bibliographystyle{IEEEtran}

\bibliography{IEEEfull,Power_Traffic_ref}

\begin{thebibliography}{10}
\providecommand{\url}[1]{#1}
\csname url@samestyle\endcsname
\providecommand{\newblock}{\relax}
\providecommand{\bibinfo}[2]{#2}
\providecommand{\BIBentrySTDinterwordspacing}{\spaceskip=0pt\relax}
\providecommand{\BIBentryALTinterwordstretchfactor}{4}
\providecommand{\BIBentryALTinterwordspacing}{\spaceskip=\fontdimen2\font plus
\BIBentryALTinterwordstretchfactor\fontdimen3\font minus
  \fontdimen4\font\relax}
\providecommand{\BIBforeignlanguage}[2]{{%
\expandafter\ifx\csname l@#1\endcsname\relax
\typeout{** WARNING: IEEEtran.bst: No hyphenation pattern has been}%
\typeout{** loaded for the language `#1'. Using the pattern for}%
\typeout{** the default language instead.}%
\else
\language=\csname l@#1\endcsname
\fi
#2}}
\providecommand{\BIBdecl}{\relax}
\BIBdecl

\bibitem{chen12g2015seeds}
C.-M. Cheng, S.-L. Tsao, and P.-Y. Lin, ``Seeds: A solar-based energy-efficient
  distributed server farm,'' \emph{IEEE Transactions on Systems, Man, and
  Cybernetics: Systems}, vol.~45, no.~1, pp. 143--156, 2015.

\bibitem{cui2015sol12ar}
M.~Cui, J.~Zhang, A.~Florita, B.-M. Hodge, D.~Ke, and Y.~Sun, ``Solar power
  ramp events detection using an optimized swinging door algorithm,'' in
  \emph{Proc. ASME Int. Design Eng. Tech. Conf. Comput. Inf. Eng. Conf}, 2015.

\bibitem{gu2016kno12wledge}
Y.~Gu, H.~Jiang, Y.~Zhang, J.~J. Zhang, T.~Gao, and E.~Muljadi, ``Knowledge
  discovery for smart grid operation, control, and situation awareness¡ªa big
  data visualization platform,'' in \emph{North American Power Symposium
  (NAPS), 2016}.\hskip 1em plus 0.5em minus 0.4em\relax IEEE, 2016, pp. 1--6.

\bibitem{katsigiann123is2010novel}
Y.~A. Katsigiannis, P.~S. Georgilakis, and G.~J. Tsinarakis, ``A novel colored
  fluid stochastic petri net simulation model for reliability evaluation of
  wind/pv/diesel small isolated power systems,'' \emph{IEEE Transactions on
  Systems, Man, and Cybernetics-Part A: Systems and Humans}, vol.~40, no.~6,
  pp. 1296--1309, 2010.

\bibitem{tian2013re1view}
Y.~Tian and C.-Y. Zhao, ``A review of solar collectors and thermal energy
  storage in solar thermal applications,'' \emph{Applied energy}, vol. 104, pp.
  538--553, 2013.

\bibitem{connolly2010review}
D.~Connolly, H.~Lund, B.~V. Mathiesen, and M.~Leahy, ``A review of computer
  tools for analysing the integration of renewable energy into various energy
  systems,'' \emph{Applied Energy}, vol.~87, no.~4, pp. 1059--1082, 2010.

\bibitem{gonzal65es2013evening}
E.~J. Gonzales and C.~F. Daganzo, ``The evening commute with cars and transit:
  Duality results and user equilibrium for the combined morning and evening
  peaks,'' \emph{Procedia-Social and Behavioral Sciences}, vol.~80, pp.
  249--265, 2013.

\bibitem{cheng2015urb1an}
Y.-H. Cheng, Y.-H. Chang, and I.~Lu, ``Urban transportation energy and carbon
  dioxide emission reduction strategies,'' \emph{Applied Energy}, vol. 157, pp.
  953--973, 2015.

\bibitem{huang2017para123meterized}
Z.~Huang, X.~Xu, H.~He, J.~Tan, and Z.~Sun, ``Parameterized batch reinforcement
  learning for longitudinal control of autonomous land vehicles,'' \emph{IEEE
  Transactions on Systems, Man, and Cybernetics: Systems}, 2017.

\bibitem{fol1ey2013impacts}
A.~Foley, B.~Tyther, P.~Calnan, and B.~{\'O}. Gallach{\'o}ir, ``Impacts of
  electric vehicle charging under electricity market operations,''
  \emph{Applied Energy}, vol. 101, pp. 93--102, 2013.

\bibitem{liu2012asse234ssment}
C.~Liu, J.~Wang, A.~Botterud, Y.~Zhou, and A.~Vyas, ``Assessment of impacts of
  {PHEV} charging patterns on wind-thermal scheduling by stochastic unit
  commitment,'' \emph{IEEE Transactions on Smart Grid}, vol.~3, no.~2, pp.
  675--683, 2012.

\bibitem{wu2012loa234d}
D.~Wu, D.~C. Aliprantis, and L.~Ying, ``Load scheduling and dispatch for
  aggregators of plug-in electric vehicles,'' \emph{IEEE Transactions on Smart
  Grid}, vol.~3, no.~1, pp. 368--376, 2012.

\bibitem{sortomme2012opt234imal}
E.~Sortomme and M.~A. El-Sharkawi, ``Optimal scheduling of vehicle-to-grid
  energy and ancillary services,'' \emph{IEEE Transactions on Smart Grid},
  vol.~3, no.~1, pp. 351--359, 2012.

\bibitem{synchro2015jiangf234s}
H.~Jiang, Y.~Zhang, J.~J. Zhang, D.~W. Gao, and E.~Muljadi,
  ``Synchrophasor-based auxiliary controller to enhance the voltage stability
  of a distribution system with high renewable energy penetration,'' \emph{IEEE
  Transactions on Smart Grid}, vol.~6, pp. 2107--2115, 2015.

\bibitem{heymans2014econo123mic}
C.~Heymans, S.~B. Walker, S.~B. Young, and M.~Fowler, ``Economic analysis of
  second use electric vehicle batteries for residential energy storage and
  load-levelling,'' \emph{Energy Policy}, vol.~71, pp. 22--30, 2014.

\bibitem{geng2017learn89ing}
X.~Geng and L.~Xie, ``Learning the lmp-load coupling from data: A support
  vector machine based approach,'' \emph{IEEE Transactions on Power Systems},
  vol.~32, no.~2, pp. 1127--1138, 2017.

\bibitem{sadeghi2014opt234imal}
P.~Sadeghi-Barzani, A.~Rajabi-Ghahnavieh, and H.~Kazemi-Karegar, ``Optimal fast
  charging station placing and sizing,'' \emph{Applied Energy}, vol. 125, pp.
  289--299, 2014.

\bibitem{he2013op234timal}
F.~He, D.~Wu, Y.~Yin, and Y.~Guan, ``Optimal deployment of public charging
  stations for plug-in hybrid electric vehicles,'' \emph{Transportation
  Research Part B: Methodological}, vol.~47, pp. 87--101, 2013.

\bibitem{wei2017opt456imal}
W.~Wei, S.~Mei, L.~Wu, M.~Shahidehpour, and Y.~Fang, ``Optimal traffic-power
  flow in urban electrified transportation networks,'' \emph{IEEE Transactions
  on Smart Grid}, vol.~8, no.~1, pp. 84--95, 2017.

\bibitem{suh20va09lnc}
B.~Subhonmesh, S.~H. Low, and K.~M. Chandy, ``Equivalence of branch flow and
  bus injection models,'' in \emph{Communication, Control, and Computing
  (Allerton), 2012 50th Annual Allerton Conference on}.\hskip 1em plus 0.5em
  minus 0.4em\relax IEEE, 2012, pp. 1893--1899.

\bibitem{st456ott2009dc}
B.~Stott, J.~Jardim, and O.~Alsa{\c{c}}, ``Dc power flow revisited,''
  \emph{IEEE Transactions on Power Systems}, vol.~24, no.~3, pp. 1290--1300,
  2009.

\bibitem{b45tzs2002optil}
A.~G. Bakirtzis, P.~N. Biskas, C.~E. Zoumas, and V.~Petridis, ``Optimal power
  flow by enhanced genetic algorithm,'' \emph{IEEE Transactions on power
  Systems}, vol.~17, no.~2, pp. 229--236, 2002.

\bibitem{yigu2014statistical1}
Y.~Gu, H.~Jiang, Y.~Zhang, and D.~W. Gao, ``Statistical scheduling of economic
  dispatch and energy reserves of hybrid power systems with high renewable
  energy penetration,'' in \emph{2014 48th Asilomar Conference on Signals,
  Systems and Computers}, 2014, pp. 530--534.

\bibitem{pe78ng2014distributed}
Q.~Peng and S.~H. Low, ``Distributed algorithm for optimal power flow on a
  radial network,'' in \emph{2014 IEEE 53rd Annual Conference on Decision and
  Control (CDC)}.\hskip 1em plus 0.5em minus 0.4em\relax IEEE, 2014, pp.
  167--172.

\bibitem{pe015dis456ed}
------, ``Distributed algorithm for optimal power flow on an unbalanced radial
  network,'' in \emph{Decision and Control (CDC), 2015 IEEE 54th Annual
  Conference on}.\hskip 1em plus 0.5em minus 0.4em\relax IEEE, 2015, pp.
  6915--6920.

\bibitem{dall2013di23stributed}
E.~Dall'Anese, H.~Zhu, and G.~B. Giannakis, ``Distributed optimal power flow
  for smart microgrids,'' \emph{IEEE Transactions on Smart Grid}, vol.~4,
  no.~3, pp. 1464--1475, 2013.

\bibitem{lam212dti123bud}
A.~Y. Lam, B.~Zhang, and N.~T. David, ``Distributed algorithms for optimal
  power flow problem,'' in \emph{Decision and Control (CDC), 2012 IEEE 51st
  Annual Conference on}.\hskip 1em plus 0.5em minus 0.4em\relax IEEE, 2012, pp.
  430--437.

\bibitem{li2012dem45and}
N.~Li, L.~Chen, and S.~H. Low, ``Demand response in radial distribution
  networks: Distributed algorithm,'' in \emph{Signals, Systems and Computers
  (ASILOMAR), 2012 Conference Record of the Forty Sixth Asilomar Conference
  on}.\hskip 1em plus 0.5em minus 0.4em\relax IEEE, 2012, pp. 1549--1553.

\bibitem{boyd2011d45istributed}
S.~Boyd, N.~Parikh, E.~Chu, B.~Peleato, and J.~Eckstein, ``Distributed
  optimization and statistical learning via the alternating direction method of
  multipliers,'' \emph{Foundations and Trends{\textregistered} in Machine
  Learning}, vol.~3, no.~1, pp. 1--122, 2011.

\bibitem{pat015fic}
M.~Patriksson, \emph{The traffic assignment problem: models and methods}.\hskip
  1em plus 0.5em minus 0.4em\relax Courier Dover Publications, 2015.

\bibitem{jia2056514nk}
N.~Jiang, C.~Xie, J.~C. Duthie, and S.~T. Waller, ``A network equilibrium
  analysis on destination, route and parking choices with mixed gasoline and
  electric vehicular flows,'' \emph{EURO Journal on Transportation and
  Logistics}, vol.~3, no.~1, pp. 55--92, 2014.

\bibitem{sheffy1985ur123ban}
Y.~Sheffy, ``Urban transportation networks: equilibrium analysis with
  mathematical programming methods,'' \emph{Traffic engineering control.
  Prentice-Hall, ISBN 0-13-93-972}, 1985.

\bibitem{he2013n1ew}
Y.~He, X.~Liu, C.~Zhang, and Z.~Chen, ``A new model for state-of-charge (soc)
  estimation for high-power li-ion batteries,'' \emph{Applied Energy}, vol.
  101, pp. 808--814, 2013.

\bibitem{pattipati2011sy11stem}
B.~Pattipati, C.~Sankavaram, and K.~Pattipati, ``System identification and
  estimation framework for pivotal automotive battery management system
  characteristics,'' \emph{IEEE Transactions on Systems, Man, and Cybernetics,
  Part C (Applications and Reviews)}, vol.~41, no.~6, pp. 869--884, 2011.

\bibitem{sed12jelmaci2017hierarchical}
H.~Sedjelmaci, S.~M. Senouci, and N.~Ansari, ``A hierarchical detection and
  response system to enhance security against lethal cyber-attacks in uav
  networks,'' \emph{IEEE Transactions on Systems, Man, and Cybernetics:
  Systems}, 2017.

\bibitem{jiang2014fault}
H.~Jiang, J.~J. Zhang, W.~Gao, and Z.~Wu, ``Fault detection, identification,
  and location in smart grid based on data-driven computational methods,''
  \emph{IEEE Transactions on Smart Grid}, vol.~5, pp. 2947 -- 2956, 2014.

\bibitem{jiang2016spatial1}
H.~Jiang, X.~Dai, W.~Gao, J.~Zhang, Y.~Zhang, and E.~Muljadi,
  ``Spatial-temporal synchrophasor data characterization and analytics in smart
  grid fault detection, identification and impact causal analysis,'' \emph{IEEE
  Transactions on Smart Grid}, vol.~7, no.~5, pp. 2525--2536, 2016.

\bibitem{ten2010cyber12security}
C.-W. Ten, G.~Manimaran, and C.-C. Liu, ``Cybersecurity for critical
  infrastructures: Attack and defense modeling,'' \emph{IEEE Transactions on
  Systems, Man, and Cybernetics-Part A: Systems and Humans}, vol.~40, no.~4,
  pp. 853--865, 2010.

\bibitem{islam2012wirel12ess}
K.~Islam, W.~Shen, and X.~Wang, ``Wireless sensor network reliability and
  security in factory automation: A survey,'' \emph{IEEE Transactions on
  Systems, Man, and Cybernetics, Part C (Applications and Reviews)}, vol.~42,
  no.~6, pp. 1243--1256, 2012.

\bibitem{geng2015da12ta}
X.~Geng and L.~Xie, ``A data-driven approach to identifying system pattern
  regions in market operations,'' in \emph{Power \& Energy Society General
  Meeting, 2015 IEEE}.\hskip 1em plus 0.5em minus 0.4em\relax IEEE, 2015, pp.
  1--5.

\bibitem{jiang2017big}
H.~Jiang, Y.~Li, Y.~Zhang, J.~J. Zhang, D.~W. Gao, E.~Muljadi, and Y.~Gu, ``Big
  data-based approach to detect, locate, and enhance the stability of an
  unplanned microgrid islanding,'' \emph{Journal of Energy Engineering}, vol.
  143, no.~5, p. 04017045, 2017.

\bibitem{park2005particle}
J.-B. Park, K.-S. Lee, J.-R. Shin, and K.~Y. Lee, ``A particle swarm
  optimization for economic dispatch with nonsmooth cost functions,''
  \emph{IEEE Transactions on Power systems}, vol.~20, no.~1, pp. 34--42, 2005.

\bibitem{boyd2004conv456ex}
S.~Boyd and L.~Vandenberghe, \emph{Convex optimization}.\hskip 1em plus 0.5em
  minus 0.4em\relax Cambridge university press, 2004.

\bibitem{li2011opt789imal}
N.~Li, L.~Chen, and S.~H. Low, ``Optimal demand response based on utility
  maximization in power networks,'' in \emph{Power and Energy Society General
  Meeting, 2011 IEEE}.\hskip 1em plus 0.5em minus 0.4em\relax IEEE, 2011, pp.
  1--8.

\bibitem{bert123sekas1989parallel}
D.~P. Bertsekas and J.~N. Tsitsiklis, \emph{Parallel and distributed
  computation: numerical methods}.\hskip 1em plus 0.5em minus 0.4em\relax
  Prentice hall Englewood Cliffs, NJ, 1989, vol.~23.

\bibitem{ja123nson1991dynamic}
B.~N. Janson, ``Dynamic traffic assignment for urban road networks,''
  \emph{Transportation Research Part B: Methodological}, vol.~25, no.~2, pp.
  143--161, 1991.

\bibitem{manual1964b123ureau}
T.~A. Manual, ``Bureau of public roads,'' \emph{US Department of Commerce},
  1964.

\bibitem{chekuri200412all}
C.~Chekuri, S.~Khanna, and F.~B. Shepherd, ``The all-or-nothing multicommodity
  flow problem,'' in \emph{Proceedings of the thirty-sixth annual ACM symposium
  on Theory of computing}.\hskip 1em plus 0.5em minus 0.4em\relax ACM, 2004,
  pp. 156--165.

\bibitem{low201454convex}
S.~H. Low, ``Convex relaxation of optimal power flow¡ªpart i: Formulations and
  equivalence,'' \emph{IEEE Transactions on Control of Network Systems},
  vol.~1, no.~1, pp. 15--27, 2014.

\bibitem{gan2014con123vex}
L.~Gan and S.~H. Low, ``Convex relaxations and linear approximation for optimal
  power flow in multiphase radial networks,'' in \emph{Power Systems
  Computation Conference (PSCC), 2014}.\hskip 1em plus 0.5em minus 0.4em\relax
  IEEE, 2014, pp. 1--9.

\bibitem{IEEE8500PDS}
\BIBentryALTinterwordspacing
Distribution test feeder. [Online]. Available:
  \url{http://ewh.ieee.org/soc/pes/dsacom/testfeeders/index.html}
\BIBentrySTDinterwordspacing

\bibitem{lam2008model123ing}
W.~H. Lam, H.~Shao, and A.~Sumalee, ``Modeling impacts of adverse weather
  conditions on a road network with uncertainties in demand and supply,''
  \emph{Transportation research part B: methodological}, vol.~42, no.~10, pp.
  890--910, 2008.

\bibitem{Duck2017c}
\BIBentryALTinterwordspacing
California {ISO} today's outlook details. [Online]. Available:
  \url{http://www.caiso.com/Pages/Today's-Outlook-Details.aspx}
\BIBentrySTDinterwordspacing

\bibitem{EIA2017132US}
\BIBentryALTinterwordspacing
Eia us energy information administration regional wholesale markets. [Online].
  Available:
  \url{https://www.eia.gov/electricity/monthly/update/wholesale_markets.php}
\BIBentrySTDinterwordspacing

\bibitem{de2011traf78fic}
A.~de~Palma and R.~Lindsey, ``Traffic congestion pricing methodologies and
  technologies,'' \emph{Transportation Research Part C: Emerging Technologies},
  vol.~19, no.~6, pp. 1377--1399, 2011.

\end{thebibliography}
\end{document}